\documentclass[aps,a4paper,showkeys,nofootinbib,longbibliography,11pt,notitlepage]{revtex4-2}

\usepackage[utf8]{inputenc}
\usepackage{filecontents}
\usepackage{natbib}
\usepackage{amsmath,amssymb,bm,amsthm}
\usepackage{subfigure}
\usepackage{graphicx}
\usepackage{dcolumn}
\usepackage{bm}
\usepackage[mathlines]{lineno}
\usepackage{booktabs}
\usepackage{color}
\newcounter{one}
\usepackage{url}

\usepackage{textcase}

\usepackage{colortbl}
\usepackage{tabularx}
\usepackage{verbatim}
\usepackage{multirow}

\usepackage[T1]{fontenc} 
\usepackage{lmodern}
\usepackage{bbm}
\usepackage[utf8]{inputenc}
\usepackage{amsfonts}
\usepackage{array}

\usepackage{bbm}
\usepackage{enumitem}
\usepackage{umoline}
\usepackage[usenames,svgnames]{xcolor}
\usepackage{natbib}
\usepackage[hyperindex,breaklinks]{hyperref}
\hypersetup{
     colorlinks=true,       		
     linkcolor=Navy,          	
     citecolor=Navy,            
     filecolor=Navy,      		
     urlcolor=Navy,           	
    runcolor=cyan,
 }
\usepackage{graphicx}
\usepackage{amsfonts}
\usepackage{amssymb}
\usepackage{amsmath}
\usepackage{bbm}
\usepackage{enumitem}

\usepackage{braket}

\newcommand{\tr}[0]{ {\rm tr}}

\newcommand{\half}[1]{{ \rm h}}

\newcommand{\for}[0]{\quad \textrm{for} \quad}

\newcommand{\dist}{d}

\newcommand{\co}{{\rm c}}

\newcommand{\cl}{{\rm cl}}

\newcommand{\Cor}{{\rm Cor}}

\def\<{\langle}
\def\>{\rangle}

\def\tr{{\rm tr}}

\newtheorem{theorem}{Theorem}
\newtheorem{subtheorem}{Subtheorem}
\newtheorem{lemma}{Lemma}
\newtheorem{corol}[lemma]{Corollary}

\newtheorem{prop}[lemma]{Proposition} 

\newcommand{\br}[1]{\left( #1 \right)}
\newcommand{\abs}[1]{\left | #1 \right|}
\newcommand{\brr}[1]{\left[ #1 \right]}
\newcommand{\brrr}[1]{\left\{ #1 \right\}}
 \newcommand{\norm}[1]{\left \|  #1 \right \|}

\newcommand{\tc}[0]{{\rm t}}

\usepackage{mathtools}
\def\multiset#1#2{\ensuremath{\left(\kern-.3em\left(\genfrac{}{}{0pt}{}{#1}{#2}\right)\kern-.3em\right)}}

\begin{document}



\title{Clustering theorem in 1D long-range interacting systems at arbitrary temperatures}

\author{Yusuke Kimura$^{1}$}
\email{yusuke.kimura.cs@riken.jp}

\author{Tomotaka Kuwahara$^{1,2,3}$}
\email{tomotaka.kuwahara@riken.jp}
\affiliation{$^{1}$ 
Analytical quantum complexity RIKEN Hakubi Research Team, RIKEN Center for Quantum Computing (RQC), Wako, Saitama 351-0198, Japan
}

\affiliation{$^{2}$ 
RIKEN Cluster for Pioneering Research (CPR), Wako, Saitama 351-0198, Japan
}
\affiliation{$^{3}$
PRESTO, Japan Science and Technology (JST), Kawaguchi, Saitama 332-0012, Japan}

\begin{abstract}

This paper delves into a fundamental aspect of quantum statistical mechanics---the absence of thermal phase transitions in one-dimensional (1D) systems. Originating from Ising's analysis of the 1D spin chain, this concept has been pivotal in understanding 1D quantum phases, especially those with finite-range interactions, as extended by Araki.  In this work, we focus on quantum long-range interactions and successfully derive a clustering theorem applicable to a wide range of interaction decays at arbitrary temperatures. This theorem applies to any interaction forms that decay faster than $r^{-2}$ and does not rely on translation invariance or infinite system size assumptions. Also, we rigorously established that the temperature dependence of the correlation length is given by $e^{{\rm const.} \beta}$, which is the same as the classical cases. Our findings indicate the absence of phase transitions in 1D systems with super-polynomially decaying interactions, thereby expanding upon previous theoretical research. To overcome significant technical challenges originating from the divergence of the imaginary-time Lieb--Robinson bound, we utilize the quantum belief propagation to refine the cluster expansion method. This approach allowed us to address divergence issues effectively and contributed to a deeper understanding of low-temperature behaviors in 1D quantum systems.

\end{abstract}

\maketitle

\tableofcontents


\section{Introduction}

One of the most fundamental aspects of quantum statistical mechanics is the absence of thermal phase transition in one-dimensional (1D) systems. This concept has historical roots tracing back to the seminal work of Ising \cite{ising2017fate}, who solved the one-dimensional spin chain problem, thereby illuminating the lack of phase transition in 1D. Building upon this, Araki~\cite{Araki1969} extended these insights into the quantum domain by investigating general quantum spin systems with finite-range interactions, particularly in the thermodynamic limit. 
In quantum equilibrium states, non-critical phases are typically characterized by a finite correlation length for bipartite correlations within quantum Gibbs states. Furthermore, the analyticity of the partition function \cite{Dobrushin1987} and the uniqueness of the Kubo--Martin--Schwinger (KMS) state \cite{doi:10.1143/JPSJ.12.570,PhysRev.115.1342,10.1143/PTPS.64.12} are also recognized as alternative hallmarks of non-criticality in these systems.

In contrast to one-dimensional (1D) systems, thermal phase transitions are more prevalent in higher-dimensional systems and have been a subject of extensive study in the history of statistical mechanics~\cite{PhysRev.65.117,stanley1971phase}. In these higher-dimensional settings, the absence of phase transitions is typically only guaranteed at higher temperatures. Interestingly, even within the confines of one dimension, it has been known that long-range interactions, characterized by a slow decay in interaction strength, can lead to thermal phase transitions.
This poses a significant question regarding the occurrence of thermal phase transitions under interactions of varying ranges, a topic that has been actively researched for many decades. The nature of these interactions is often modeled by the polynomial decay with distance $r$, expressed as $r^{-\alpha}$. Here, the critical value of $\alpha$ plays a crucial role: In classical theory, if $\alpha \leq 2$, it indicates the presence of a thermal phase transition~\cite{Dyson1969,PhysRevLett.29.917,PhysRevLett.37.1577,PhysRev.187.732}, whereas $\alpha > 2$ suggests its absence \cite{Dobrushin1973}. Further insights can be found in the footnote~\footnote{Precisely, the condition is related to the finiteness of the boundary energy between two split subsystems. Generally, for polynomially decaying interactions, this finiteness is assured when $\alpha > 2$~\cite{kuwahara2019area}.}.

A pivotal question in the study of quantum long-range interacting chains is the determination of the regime of $\alpha$ that ensures non-criticality at arbitrary temperatures. While initial studies, such as those by Araki~\cite{Araki1975} and Kishimoto~\cite{Kishimoto1976}, have established the uniqueness of the KMS state in 1D systems with long-range interactions for $\alpha > 2$, 
it has been still challenging to characterize the 1D quantum phases beyond these foundational works. Remarkably, it has been over half a century since the generalization of the methodologies used in Araki's seminal work~\cite{Araki1969}.
In \cite{Araki1969}, the analysis of the 1D Ising model was conducted using an intricate operator algebra technique. This advanced mathematical approach enabled the exploration of phase transitions within a specific scope, notably in systems where interactions were of finite range and exhibited translational invariance. However, the complexity of these methods presented substantial challenges in generalizing the analyses to broader system classes, particularly in contexts where these constraints were not met. As a result, the pursuit of understanding phase transitions in more generalized settings remained a formidable task, underscoring the need for novel approaches in this field of research.

In recent developments, the paper by P\'erez-Garc\'ia et al.~\cite{Perez-Garcia2023} has shed light on the challenges in generalizing the understanding of quantum long-range interacting systems. A significant hurdle identified is the divergence of the imaginary-time Lieb--Robinson bound, which describes the quasi-locality of systems after undergoing imaginary-time evolution. This divergence, as generally proven in \cite{10.1063/1.4936209}, implies that an imaginary-time-evolved local operator becomes entirely non-local across the whole system after a threshold time. 
However, in one-dimensional systems with finite-range interactions, this quasi-locality is preserved for any imaginary time. In these cases, the length scale of the quasi-locality increases exponentially with imaginary time, as detailed in \cite[Lemma~1]{kuwahara2018polynomial} for example. Extending beyond finite-range interactions, P\'erez-Garc\'ia et al.~\cite{Perez-Garcia2023} demonstrated that the convergence of the imaginary Lieb--Robinson bound is attainable as long as the interaction decays super-exponentially with distance. 
For interactions that decay exponentially or slower, the imaginary Lieb--Robinson bound remains convergent only below a specific threshold time.
This insight indicates that any attempt to generalize beyond exponentially decaying interactions requires the development of alternative mathematical techniques to effectively handle one-dimensional quantum Gibbs states.

In our current work, we address a portion of this open problem and derive a clustering theorem applicable to any form of interaction decay at arbitrary temperatures. The merits of our results are summarized as follows:
\begin{enumerate}
    \item The theorem is applicable to all forms of interaction, provided the decay rate is faster than $r^{-2}$ [see the assumption~\eqref{interaction_decay_J}].
    \item Our result is established unconditionally, meaning we do not assume translation invariance or infinite system size. 
    Moreover, for slower decay of interactions than subexponential form, the obtained decay rate for the bi-correlation is qualitatively optimal in the sense that it shows qualitatively the same behavior as the interaction decay (see Theorem~\ref{main_thm_clustering}).
    \item We determine the explicit form of the correlation length as a function of the inverse temperature. In detail, we identified that the correlation length is proportional to $e^{{\rm const.} \times \beta}$. This is qualitatively optimal since the correlation length in the classical Ising model has an exponential dependence on the inverse temperature. 
\end{enumerate}
These strengths enable the broad application of our findings, extending and generalizing the scope of previous works based on Araki's foundational studies \cite{Araki1969}. Examples include clustering theorems for mutual information~\cite{bluhm2021exponential} and conditional mutual information~\cite{kato2016quantum}, refined characterizations of locality properties~\cite{capel2024decay,bluhm2024strong}, 
studies on rapid thermalization~\cite{PhysRevLett.130.060401}, and the equivalence of statistical mechanical ensembles~\cite{fern2015equivalence,PhysRevLett.124.200604,kuwahara2019gaussian}, efficient simulation of the quantum Gibbs state~\cite{Fawzi2023subpolynomialtime,trivedi2023quantum}, learning the thermal states~\cite{onorati2023efficient}, among others.
However, our results are not without limitations:
\begin{enumerate}
    \item In the case of exponentially decaying interactions, the decay rate of the bipartite correlation function is subexponential, following $e^{-\Omega(\sqrt{R})}$ (where $R$ denotes distance), rather than an exponential form.
    \item Our findings do not extend to the analyticity of the partition function. For finite systems, analyzing the positions of complex zeroes in the partition function is essential~\cite{Mehdi2020}. Currently, our techniques appear limited in analyzing quantum Gibbs states at complex temperatures.
\end{enumerate}
Despite these limitations, our research significantly enhances the understanding of the structure of 1D quantum Gibbs states at low temperatures. Notably, our results suggest the absence of phase transitions in any one-dimensional system with super-polynomially decaying interactions, a significant extension from previous results \cite{Perez-Garcia2023} limited to super-exponentially decaying interactions\footnote{There is still potential for a phase transition from purely exponential decay to subexponential decay of bipartite correlations.}.

Finally, we highlight our principal technical contributions in this work. To address the challenge of the divergence of the imaginary Lieb--Robinson bound, we have integrated the quantum belief propagation technique~\cite{PhysRevB.76.201102,PhysRevB.86.245116} with the established Lieb--Robinson bound \cite{ref:LR-bound72,PhysRevLett.97.050401}. This combination is pivotal in the recent advancements in the field of low-temperature quantum many-body physics \cite{Brandao2019,Anshu_2021,PhysRevX.11.011047,PhysRevX.12.021022}.
We show the basic techniques in Sec.~\ref{sec:Basic locality techniques}. 
Our approach utilizes the quantum belief propagation method to refine the cluster expansion technique (see Sec.~\ref{sec:Cluster expansion technique}). Typically, the expansion encounters divergence beyond a certain temperature threshold, even in one-dimensional systems~\cite{Gross1979,Park1995,ueltschi2004cluster,PhysRevX.4.031019,frohlich2015some,CMI_clustering}. To circumvent this issue at arbitrary temperatures, we draw upon strategies from classical cases, where divergence can be more readily avoided through the use of a standard transfer-matrix method (see Sec.~\ref{1D cases: commuting Hamiltonian}).
In our exploration, we identify several challenges in extending these classical methods to quantum scenarios. 
Our paper proposes solutions to these challenges (see Sec.~\ref{sec:Proof of the main theorem}), primarily through careful treatment of the cluster expansion. This approach not only addresses the immediate issue of divergence but also contributes to a broader understanding of quantum systems and their behaviors at low temperatures.

\section{Setup}

\label{sec2}
Consider a quantum system consisting of $n$ spins, or qudits, arranged in a one-dimensional (1D) chain. We denote the total set of spins by $\Lambda$. 
For any arbitrary subset $X$ within $\Lambda$, denoted as $X \subseteq \Lambda$, the cardinality of $X$---the number of sites in $X$---is represented by $|X|$. Given that the system contains $n$ spins, it follows that $|\Lambda| = n$.
For a sequence of subsets $X_1,X_2,\ldots, X_q$, we use the notation $X_{1:q}$ to represent the union of the $X_i$'s, i.e., $X_{1:q}:=X_1\cup X_2 \cup \cdots \cup X_q$.   
For any two subsets $X$ and $Y$ in $\Lambda$, we define the distance between them, $\dist_{X,Y}$, as the shortest path length on the graph connecting $X$ and $Y$. It is important to note that if $X$ and $Y$ have a non-empty intersection, the distance between them is considered to be zero. Additionally, when $X$ consists of a single element, i.e., $X = \{i\}$, we simplify the notation for distance from $\dist_{\{i\},Y}$ to $\dist_{i,Y}$ for convenience.
We denote the complement of $X$ by $X^\co$, i.e., $X^\co := \Lambda \setminus X$, and the boundary of $X$ by $\partial X$, defined as $\partial X := \{ i \in X \,|\, \dist_{i,X^\co} = 1 \}$. 

We consider a general $k$-local Hamiltonian $H$, described by
\begin{align}
\label{h_Z_positive}
    H = \sum_{Z: |Z| \leq k} h_Z ,
\end{align}
where $h_Z$ is an arbitrary operator supported on the subset $Z \subset \Lambda$ with $|Z| \leq k$. 
For an arbitrary subset $L$, we denote the subset Hamiltonian supported on $L$ by $H_L$:
\begin{align}
\label{subset_Hamiltonian}
    H_L = \sum_{Z: Z\subseteq L} h_Z.
\end{align}
We characterize the interaction decay as follows: Let $g > 0$, and $\bar{J}: \mathbb{N} \to [0,\infty)$ be a monotonically decreasing function with $\bar{J}(0)=1$ such that
\begin{align}
    \label{def_short_range_long_range}
    J_{i,i'} := \sum_{Z: Z \ni \{i,i'\}} \norm{h_Z} \leq g \bar{J}(\dist_{i,i'}) .
\end{align}
Since the total system $\Lambda$ is finite, such $g$ and $\bar{J}$ always exist.
We note that the one-site energy is upper-bounded by $g$: 
\begin{align}
    \label{g_extensive}
 \sum_{Z: Z \ni i} \norm{h_Z} \leq g \bar{J}(0) = g .
\end{align}
We assume that $\bar{J}(x)$ decays faster than $1/x^2$ in the following sense: We assume that there exists $\gamma>0$ such that for $z\in \{0,1\}$ and all $\ell\in\mathbb{N}$, the following inequality holds:
\begin{align}
\label{interaction_decay_J}
\sum_{x=\ell}^\infty x^z \bar{J}(x) \le \gamma \ell^{z+1} \bar{J}(\ell).
\end{align}
For Hamiltonians with finite-range interactions, we define the term interaction length, $d_H$, such that for any distance $d_{i,j} $ larger than $d_H$ ($d_{i,j}>d_H$), $\bar{J}(d_{i,j})$ vanishes, i.e., $\bar{J}(d_{i,j})=0$.

We define the quantum Gibbs state $\rho_\beta$ at a fixed inverse temperature $\beta$ as follows:
\begin{align}
    \rho_\beta := \frac{e^{\beta H}}{Z_\beta}, \quad Z_\beta = \tr \left( e^{\beta H} \right).
\end{align}
In this definition, the minus sign appearing in the standard Gibbs state is included in the Hamiltonian itself. Throughout the analyses, we assume that the interaction terms satisfy the following condition:  
 \begin{align}
h_Z \succeq 0 \for Z\subset \Lambda.
\end{align} 
This assumption can always be satisfied by appropriately shifting the energy origin,specifically by adding a constant operator $\norm{h_Z}\hat{1}_Z$ to $-h_Z$.

In the following, $\Theta(x)$ will always denote a function of the form
\begin{align}
    \label{Theta_notation_def}
     \theta_0+\theta_1\, x,
\end{align}
with positive coefficients $\theta_0, \theta_1$ that only depend on the fundamental parameters $g$, $k$ and $\gamma$.

For an arbitrary operator $O$, we define the time evolution of $O$ by another operator $O'$ (e.g., a subset Hamiltonian $H_L$) as 
\begin{align}
\label{notatation_label_O_O',t}
O(O',t) := e^{iO't} O e^{-iO't} .
\end{align}
For simplicity, we usually denote $O(H,t)$ by $O(t)$. 

We use $O_X$ to represent an operator that acts on the Hilbert space supported on the subset $X\subset\Lambda$. Given operators $O_X$ and $O_Y$ supported on subsets $X, Y\subset\Lambda$, the correlation function $\Cor_{\rho_\beta}(O_X, O_Y)$ is defined as follows:
\begin{align}
\label{def:cor_O_X_O_Y}
\Cor_{\rho_\beta}(O_X, O_Y):=\tr (\rho_\beta O_X O_Y)-\tr (\rho_\beta O_X)\tr (\rho_\beta O_Y). 
\end{align}
The correlation length $\xi_{\beta,\delta}$ is such that for any sets $X$ and $Y$ with a distance $d_{X,Y}$ larger than $\xi_{\beta,\delta}$, i.e., $d_{X,Y}>\xi_{\beta,\delta}$, the correlation between them becomes smaller than $\delta$, where $\delta$ is an arbitrary constant. For example, when the correlation function $\Cor_{\rho_\beta}(O_X, O_Y)$ is proportional to $(d_{X,Y}/\eta_\beta)^{-\alpha}$, i.e., $\Cor_{\rho_\beta}(O_X, O_Y)=c (d_{X,Y}/\eta)^{-\alpha}$, the parameter $\eta_\beta$ characterizes the correlation length, meaning $\xi_{\beta,\delta}=\eta_\beta (c/\delta)^{1/\alpha}$. Our purpose is to identify the $\beta$ dependence of $\xi_{\beta,\delta}$ when $\delta$ is an arbitrary constant independent of $\beta$.  

For $\Lambda_0$ and $L, L'\subset \Lambda$, we define the interaction operator $V_{L,L'}(\Lambda_0)$ between $L$ and $L'$ as follows:
\begin{align}
V_{L,L'}(\Lambda_0) :=\sum_{\substack{Z: Z\subset \Lambda_0 \\ Z\cap L\neq \emptyset , Z\cap L'\neq \emptyset} }h_Z  .
\label{Def:V_X_Y}
\end{align} 
$V_{L,L'}(\Lambda_0)$ characterizes the interactions between blocks $L$ and $L'$ acting on the region $\Lambda_0$. Following Ref.~\cite[Supplementary Lemma~2]{kuwahara2019area}, we prove the following statement:
\begin{lemma} \label{lem_block_int}
The norm of $V_{L,L'}(\Lambda_0)$ is bounded above by
 \begin{align}
\|V_{L,L'}(\Lambda_0)\| \le g \gamma^2 r^2 \bar{J}(r) ,  \label{definition_of_H_X_l0}
\end{align}
where $r$ represents the distance between $L$ and $L'$, and the function $\bar{J}(r)$ is defined in Eq.~\eqref{def_short_range_long_range}. 
\end{lemma}

\textit{Proof of Lemma~\ref{lem_block_int}.}
In our proof, we aim to calculate an upper bound for the term $\overline{V}_{L,L'}$, defined as:
\begin{align}
    \overline{V}_{L,L'} := \sum_{\substack{Z:Z\cap L \neq \emptyset, \\ Z\cap L' \neq \emptyset}} \| h_Z\|, \label{upper_bound_overline_H_X_l0}
\end{align}
which serves as an upper bound on $\|V_{L,L'}(\Lambda_0)\|$ for any subset $\Lambda_0 \subset \Lambda$.
Assuming that subsystem $L'$ is located to the right of $L$, we define $L=\{i_0-n_L+1, i_0-n_L+2, \ldots, i_0\}$, and $L' = \{i_0+r, i_0+r+1, \ldots, i_0+r+n_{L'}-1 \}$. We denote the cardinalities of $L$ and $L'$ by $n_L$ and $n_{L'}$, respectively, i.e., $n_L:=|L|$ and $n_{L'} := |L'|$. Using the equality $\dist(i_0+j, L) = r-1+j$, we obtain:
\begin{align}
    \sum_{\substack{Z:Z\cap L \neq \emptyset, \\ Z\cap L' \neq \emptyset}} \| h_Z\| 
    &\leq \sum_{j=0}^{n_{L}-1}\sum_{j'=0}^{n_{L'}-1}  \sum_{Z:Z\ni  \{ i_0-j, i_0+r+j'\}} \|h_Z\| 
    \leq g \sum_{j=0}^{n_{L}-1}\sum_{j'=0}^{n_{L'}-1}  \bar{J}(r+j+j') ,
    \label{Interaction_between_X_Y_L_1}
\end{align}
where we used inequality~\eqref{def_short_range_long_range}.
Furthermore, substituting $z=0$ and $\ell=r+j$ into inequality~\eqref{interaction_decay_J} yields the following:  
\begin{align}
\label{ineq1}
\sum_{j'=0}^{n_{L'}-1}  \bar{J}(r+j+j')  \le \sum_{x=r+j}^{\infty}  \bar{J}(x) \le \gamma (r+j)\bar{J}(r+j) .
\end{align}
Similarly, substituting $z=1$ and $\ell=r$ into inequality~\eqref{interaction_decay_J}, we obtain: 
\begin{align}
\label{ineq2}
\sum_{j=0}^{n_{L}-1}(r+j)\bar{J}(r+j) \le \sum_{x=r}^{\infty}x\bar{J}(x) \le  \gamma r^2\bar{J}(r) .
\end{align}
Applying inequalities \eqref{ineq1} and \eqref{ineq2} to~\eqref{Interaction_between_X_Y_L_1} results in the inequality: 
\begin{align}
    \overline{V}_{L,L'} =    \sum_{\substack{Z:Z\cap L \neq \emptyset, \\ Z\cap L' \neq \emptyset}} \| h_Z\|     &\leq g \gamma^2 r^2 \bar{J}(r) ,
\end{align}
which concludes the proof. $\square$

\subsection{Main results}
For our purpose, we categorize the forms of interaction into two classes: i) faster than or equal to sub-exponential decay, i.e., $\bar{J}(r) \propto e^{-r^{\kappa}}$, $(0 < \kappa \le 1)$~\footnote{We can treat the super-exponentially decaying interactions by reducing them to the case of $\kappa=1$. 
The decay rate of the Lieb--Robinson bound cannot be improved from the exponential form since the condition~\eqref{cond:Infinite interaction length} breaks down for $\bar{J}(r) \propto e^{-r^{\kappa}}$ with $\kappa>1$. }, and ii) slower than sub-exponential decay, i.e., $\lim_{r\to \infty}e^{-r^\kappa}/\bar{J}(r) \to 0$ for $\forall \kappa>0$. 
For both types of interaction decay, we can prove the following theorem:
\begin{theorem} \label{main_thm_clustering}
Let $O_X$ and $O_Y$ be arbitrary unit-norm operators supported on subsets $X$ and $Y$ ($X,Y\subset \Lambda$), $\norm{O_X}=\norm{O_Y}=1$, and $R:=d_{X,Y}$. If there exist $\kappa>0$ and $c>0$ such that $\bar{J}(r) \le e^{-cr^\kappa}$ for all $r\ge 0$ (i.e., when the interaction decays subexponentially or faster), then the correlation function $\Cor_{\rho_\beta}(O_X,O_Y)$ satisfies the following inequality:
\begin{align}
\label{main:clustering_rho_subexp}
&\Cor_{\rho_\beta}(O_X,O_Y)  \le c_1 e^{-R_\beta^{\kappa/(\kappa+1)}} 
\end{align}
for some positive real constant $c_1$ that depends only on the system details, where $R_\beta = R/\xi_\beta$ and $\xi_\beta := e^{\Theta (\beta)}$. $\Theta(\beta)$ is a function of the form \eqref{Theta_notation_def}.

In the case where the interaction decays more slowly than subexponential forms, there exists a positive real constant $c_2$, and the following inequality holds:
 \begin{align}
\label{main:clustering_rho_slower}
\Cor_{\rho_\beta}(O_X,O_Y)  
&\le c_2 R_\beta^2  \bar{J}\left(R_\beta/\log^2\left(\bar{J}(R)^{-1}\right)\right).
\end{align}
For any $\varepsilon>0$ there exists $r_{0,\beta}$ such that for $R>r_{0,\beta}$,
\begin{align}
\bar{J}\left(R_\beta/\log^2\left(\bar{J}(R)^{-1}\right)\right) \le \bar{J}\left(R_\beta^{1-\epsilon}\right).
\end{align}
\end{theorem}

{\bf Remark.} We mention several points below:
\begin{enumerate}

\item{} The bound in \eqref{main:clustering_rho_subexp} is independent of $|X|$ and $|Y|$. For practical applications, it is quite important to keep track of how the bound depends on the size of the supports. See \cite{capel2024decay} for relevant discussion. 

\item{} In deriving the results, we do not need to assume translation invariance or an infinite system size. 
The key technique we use is quantum belief propagation (discussed in Sec.~\ref{sec::Quantum belief propagation}), along with the Lieb--Robinson bound (to be discussed in Sec.~\ref{sec2.2}), as an alternative to the imaginary Lieb--Robinson bound, which played a crucial role in the original works~\cite{Perez-Garcia2023,Araki1969}. 

\item{} We identified the dependence of the correlation length on the inverse temperature, $\beta$. 
The correlation length $\xi_{\beta,\delta}$ defined below Eq.~\eqref{def:cor_O_X_O_Y} is now given by 
\begin{align}
\xi_{\beta,\delta} = K(\delta) \xi_\beta 
\end{align}
with $K: \mathbb{R}^+ \to [0,\infty)$, where the explicit form of $K(\delta)$ depends on the interaction decay $\bar{J}(r)$. 
From the above expression, it can be seen that $\beta$ dependence of the correlation length is at most exponential.
As far as we know, this is the first result to rigorously prove that the correlation length follows the form $\xi_\beta=e^{{\rm const} \times \beta}$. 
This type of dependence is often assumed as empirical knowledge without proof, e.g., in Ref.~\cite{kato2016quantum}.

\item{} When the interaction decay is polynomial, such as $\bar{J}(r) \propto r^{-\alpha}$, the clustering theorem states that there is a positive real constant $c_3$ and the inequality 
  \begin{align}
\label{main:clustering_rho_long-range}
\Cor_{\rho_\beta}(O_X,O_Y)  
&\le  c_3 R_\beta^{-\alpha+2} \log^{2\alpha}(R) = \tilde{\mathcal{O}}(R^{-\alpha+2}) 
\end{align}
holds.

\item{} The bipartite correlation is expected to be lower-bounded by the interaction strength between $X$ and $Y$:
\begin{align}
&\Cor_{\rho_\beta}(O_X,O_Y)  \gtrsim \norm{V_{X,Y}(\Lambda)} \propto R^2 \bar{J}(R) .
\end{align}
The latter relation is obtained when we choose $X=(-\infty, x]$ and $Y=[x+R+1,\infty)$ in the thermodynamic limit. 
Therefore, the clustering theorem~\eqref{main:clustering_rho_slower} for slower-decaying interactions is expected to be qualitatively optimal in the limit at which $|X| \to \infty$ and $|Y| \to \infty$. 

\item{} On the other hand, when there exist small positive constants $n_1$ and $n_2$, such that $|X|\le n_1$ and $|Y|\le n_2$ hold, the upper bound may be improved to the following form: 
\begin{align}
&\Cor_{\rho_\beta}(O_X,O_Y)  \lesssim \norm{V_{X,Y}(\Lambda)} \propto |X| \cdot |Y| \bar{J}(R) . 
\end{align}
At this stage, our current analyses cannot be straightforwardly extended to derive this upper bound. 
For example, at high temperatures, one can prove an improved version of the clustering theorem in long-range interacting systems~\cite{Kim-Kuwahara-Saito}.
In this context, the optimal clustering theorem for such $|X|$ and $|Y|$ remains open, even for power-law interaction decays.

\item{} For sub-exponentially decaying interactions, our results are far from optimal. In particular, for $\kappa=1$, our bound gives 
\begin{align}
\label{main:clustering_rho_subexp___2}
&\Cor_{\rho_\beta}(O_X,O_Y) = e^{-\Omega\br{\sqrt{R_\beta}}}  \for \bar{J}(r) \le e^{-{\rm const.} r} ,
\end{align}
which is worse than the previous results in Refs.~\cite{Perez-Garcia2023,Araki1969}, which achieved exponential decay of correlation for arbitrary interaction decay with $\kappa>1$.  
Still, our result has the advantage of not requiring threshold temperatures as in Ref.~\cite{Perez-Garcia2023}, and our theorem holds without assuming an infinite system size or translation invariance.

\end{enumerate}

\subsection{Technical overview for the proof}

The main techniques are summarized in the following three points:
\begin{enumerate}
\item{} Construction of the interaction-truncated Hamiltonian (Sec. \ref{sec:Interaction-truncated Hamiltonian}), 
\item{} Cluster expansion technique (Sec.~\ref{sec:Cluster expansion technique}), 
\item{} Efficient block decomposition using quantum belief propagation operators (Sec.~\ref{sec:Decomposition of the system}). 
\end{enumerate}
The first technique serves as the foundation for previous work on proving the long-range entanglement area law~\cite{kuwahara2019area}. 
Here, we do \textit{not} simply truncate all the long-range interactions across the entire region, but only those around the boundary of interest (in this case, the region between $X$ and $Y$. See also Fig.~\ref{Fig:Interaction_Truncate} below). 
For this interaction truncated Hamiltonian $H_{\rm t}$, the Lieb--Robinson bound reads [Corollary~\ref{LR_Lieb_Robinson:short} below]
 \begin{align}
\norm{[O_{Z}(H_{\rm t}, t), O_{Z'}]}\le \norm{O_{Z}} \cdot \norm{O_{Z'}} \cdot |Z| \cdot |Z'| e^{v|t|} \tilde{\mathcal{F}}(r) , \quad r=\dist_{Z,Z'}, 
\end{align}
where $\tilde{\mathcal{F}}(r)$ is defined as $\tilde{\mathcal{F}}(r) = C \min \br{ e^{-r/(2l_0)} , \bar{J}(r)}$ with $C$ a constant defined in Eq.~\eqref{Lieb_robinson_func} and $l_0$ is the cutoff interaction length. 
Therefore, under interaction truncation, exponential quasi-locality through time evolution is restored.    
Additionally, in inequality~\eqref{truncation:error_norm}, we can show that the closeness between $e^{\beta H}$ and $e^{\beta H_{\rm t}}$ is approximately estimated as
$$\norm{e^{\beta H} - e^{\beta H_{\rm t}}}_1  \le \beta R l_0 \bar{J}(l_0) Z_\beta$$ with $\bar{J}(l_0)$ representing the decay rate of interaction as in~\eqref{def_short_range_long_range}.

As the second analytical tool, we employ the cluster expansion technique~\cite{PhysRevX.4.031019,frohlich2015some}. 
Following the notation in Ref.~\cite{PhysRevX.4.031019} (see also Sec.~\ref{sec:Cluster expansion technique}), we express the correlation function as ${\rm Cor}_{\rho_\beta}(O_X,O_Y)=\tr(\rho_{\rm cl} O_X^{(0)} O_Y^{(1)})$, where $\rho_{\rm cl}$ is given by the sum of the relevant cluster terms [which will be presented later in Eq.~\eqref{upper_bound_beta_m}]:
\begin{align}
\label{correlation_expansion}
\rho_{\rm cl}:=\frac{1}{Z_\beta^2 } \sum_{m=0}^\infty \sum_{Z_1,Z_2,\ldots,Z_m: {\rm connected}} \frac{\beta^m}{m!} h_{Z_1}^{(+)} h_{Z_2}^{(+)} \cdots h_{Z_m}^{(+)}. 
\end{align}
Here, the operators $O^{(+)}$, $O^{(0)}$ and $O^{(1)}$ are defined by applying a copy of the original Hilbert space: 
$O^{(+)} = O \otimes \hat{1} +  \hat{1} \otimes O$, $O^{(0)} = O \otimes \hat{1}$ and $O^{(1)} =  O \otimes \hat{1} -  \hat{1} \otimes O$.
As is well-known~\cite{Kotecky1986,PRXQuantum.5.010305}, each term's norm in the expansion~\eqref{correlation_expansion} is bounded above by the form $ \sum_{m=0}^\infty(\beta/\beta_c)^m$, and this upper bound diverges for $\beta >\beta_c$. 
Because of this, even in one-dimensional cases, the method of simply counting the connected clusters does not work for values of $\beta$ above a threshold $\beta_c$. 
In 1D cases, the cluster expansion~\eqref{correlation_expansion} corresponds to the removal of zeroth-order interaction terms that connect subsets $X$ and $Y$. 
In classical (i.e., commuting) cases, this method allows us to efficiently count connected clusters, analogous to the transfer-matrix method (Sec.~\ref{1D cases: commuting Hamiltonian} below).

When extending the analysis from commuting to non-commuting cases, we encounter a significant challenge: we cannot independently eliminate the zeroth-order interaction terms. This difficulty arises because operators do not commute in the non-commuting case. 
To explain this issue in more detail, consider the exponential operator $e^{\lambda_1 O_1 + \lambda_2 O_2}$ and the task of removing the zeroth-order terms for $\lambda_1$ and $\lambda_2$.
In the commuting case, the zeroth-order terms can be easily removed by factorizing as follows:
$$\br{e^{\lambda_1 O_1} -1} \br{ e^ {\lambda_2 O_2}-1}.$$ 
However, in the non-commuting case, this factorization is not possible because $[O_1,O_2]\neq 0$. 
A physical interpretation of this is the following. In general, the effects of local interactions for quantum Gibbs states are not confined to a specific location but spread over distant sites.

To address this challenge, we partition the entire system into blocks of length $2\ell$, in such a way that, approximately, the central sites of the resulting blocks are independent of one another, as illustrated in Fig.~\ref{fig:L_s_Phi_s}. 
The approximation error, $\mathcal{G}_\beta(\ell)$, is evaluated based on the quasi-local nature of quantum belief propagation, as detailed in Proposition~\ref{lem:belief_error}. 
This evaluation can also be done using imaginary time evolution, provided that the interaction decay is faster than exponential. However, quantum belief propagation offers better error estimates and applies to all forms of interaction. 
For each central site of the blocks, it becomes possible to effectively eliminate the zeroth-order terms, allowing us to approximate the correlation decay rate as $e^{-m/e^{\Theta(\beta)}}$, where $m$ represents the number of blocks, which scales with $R/\ell$ (see also Subtheorem~\ref{sbthm:tildeG_upp}). 
Conversely, to reduce the approximation error $\mathcal{G}_\beta(\ell)$, a larger block size $\ell$ is preferred. However, choosing a larger block size reduces the number of blocks, $m\propto R/\ell$, and thus weakens the decay rate to $e^{-m/e^{\Theta(\beta)}}$. Selecting optimal values for $m$ and $\ell$ to minimize both $\mathcal{G}_\beta(\ell)$ and $e^{-m/e^{\Theta(\beta)}}$ is crucial for our main theorem, with explicit choices shown in Sec.~\ref{seec/:Completing the proof}. 
We note that this trade-off between $\mathcal{G}_\beta(\ell)$ and $e^{-m/e^{\Theta(\beta)}}$ prevents us from achieving the optimal clustering theorem for interactions that decay faster than sub-exponentially (see also the concluding remarks in Section~\ref{sec5}).

\section{Basic locality techniques} \label{sec:Basic locality techniques}

\subsection{Lieb--Robinson bound}
\label{sec2.2}
The Lieb--Robinson bound captures the quasilocality of dynamics through time evolution \cite{ref:LR-bound72, ref:Hastings2006-ExpDec, Nachtergaele2006, ref:Nachtergaele2006-LR} by providing a bound on the error when approximating the dynamics with a light cone. This bound plays a key role in deriving the main result of this work, and we briefly review its formulation here. We first discuss the case where the Hamiltonian has a finite interaction length. (Recall that we introduced the term "interaction length" below equation \eqref{g_extensive}.)  

\begin{lemma}[Finite-range interaction]
\label{lemma_2}
Let $H$ be a Hamiltonian with interaction length $d_H$. Then, for any operators $O_{Z}$ and $O_{Z'}$ with $d_{Z,Z'}=r$, the following relation holds for the commutator $[O_{Z}(t), O_{Z'}]$:
\begin{align}
\label{LR_bnd_finite}
\norm{[O_{Z}(t), O_{Z'}]}\le \frac{2}{k} \norm{O_{Z}} \cdot \norm{O_{Z'}} \cdot |Z|   \frac{(2gk|t|)^{n_0}}{n_0!},
\end{align}
where the number $n_0$ is defined by
\begin{align}
n_0=\left \lfloor \frac{r}{d_H} +1 \right \rfloor.
\end{align} 
\end{lemma}

Here, $\left \lfloor \cdot \right \rfloor$ represents rounding down. By definition, $n_0$ is larger than $\frac{r}{d_H}$, i.e., $n_0 > \frac{r}{d_H}$. A proof of Lemma \ref{lemma_2} can be found, for example, in~\cite[Ineq. (24)]{Phd_kuwahara}. 

{\bf Remark.} 
Using the following inequality, which holds for $x, x_0>0$:
\begin{align}
\br{\frac{x_0}{x}}^x \le e^{-x+x_0},
\end{align}
inequality~\eqref{LR_bnd_finite} can be rewritten as 
\begin{align}
\label{LR_bnd_finite__2}
\norm{[O_{Z}(t), O_{Z'}]}\le \frac{2}{k} \norm{O_{Z}} \cdot \norm{O_{Z'}} \cdot |Z| e^{-r/d_H + 2egk|t|},
\end{align}
where we used the inequality $n!\ge (n/e)^n$.

Next, we consider the case where the Hamiltonian has an infinite interaction length (i.e., when $d_H=\infty$). The following lemma, whose application is central to the subsequent discussion, applies to this case:
\begin{lemma}[Infinite interaction length]
\label{lemma_3}
Let $H$ be a Hamiltonian with an infinite interaction length. 
The parameter $\mathfrak{g}$ is defined as the infimum over all constants $c$ that satisfy the inequality 
\begin{align}
\label{cond:Infinite interaction length}
\max_{i,i'\in \Lambda} \sum_{i_0\in \Lambda} \bar{J}(\dist_{i,i_0}) \bar{J}(\dist_{i_0,i'})\le c \bar{J}(\dist_{i,i'}).
 \end{align}
The set of such constants is nonempty (and therefore, $\mathfrak{g}$ is well-defined). The Lieb--Robinson bound for the commutator $[O_{Z}(t), O_{Z'}]$ is then given by
\begin{align}
\label{LR_bnd_infinite__infinite}
\norm{[O_{Z}(t), O_{Z'}]}\le \frac{2}{\mathfrak{g}} \norm{O_{Z}} \cdot \norm{O_{Z'}} \cdot |Z| \cdot |Z'| \br{ e^{2\mathfrak{g}|t|}-1} \bar{J}(r) ,
\end{align}
where $r:= \dist_{Z,Z'}$.  
\end{lemma}
A proof of Lemma \ref{lemma_3} can be found, e.g., in~\cite[Theorem A.1]{ref:Hastings2006-ExpDec}.

Combining the two Lieb--Robinson bounds mentioned above provides an upper bound on the norm of the commutator $[O_{Z}(t), O_{Z'}]$ as follows: 
\begin{align}
\label{bound_comm_cor}
\norm{[O_{Z}(t), O_{Z'}]}
&\le \norm{O_{Z}} \cdot \norm{O_{Z'}} \cdot |Z| \cdot |Z'| \min\brr{2, e^{v|t|} \mathcal{F}_0(r) } \notag \\
&=: \norm{O_{Z}} \cdot \norm{O_{Z'}} \cdot |Z| \cdot |Z'| \mathcal{F}(t,r) ,
\end{align}
with
\begin{align}
\label{def:mathcal_F_(r)}
\mathcal{F}(t,r) :=  \min\brr{2, e^{v|t|} \mathcal{F}_0(r)},
\end{align}
and 
\begin{align}
\label{Lieb_robinson_func}
\mathcal{F}_0(r) = \begin{cases} 
Ce^{-r/d_H} &\for \textrm{finite-range interactions}, \\ 
 C \bar{J}(r) &\for \textrm{infinite-range interactions} ,\\ 
 \end{cases}
\end{align}
$C=2(1/k+1/\mathfrak{g})$, and $v=\max(2gk,2\mathfrak{g})$. 
The norm of the commutator $\norm{[O_{Z}(t), O_{Z'}]}$ also has the following trivial bound: $\norm{[O_{Z}(t), O_{Z'}]}\le 2\norm{O_{Z}} \cdot \norm{O_{Z'}} $. In subsequent discussions, we will utilize this trivial upper bound.

\begin{figure}
\centering
\includegraphics[clip, scale=0.4]{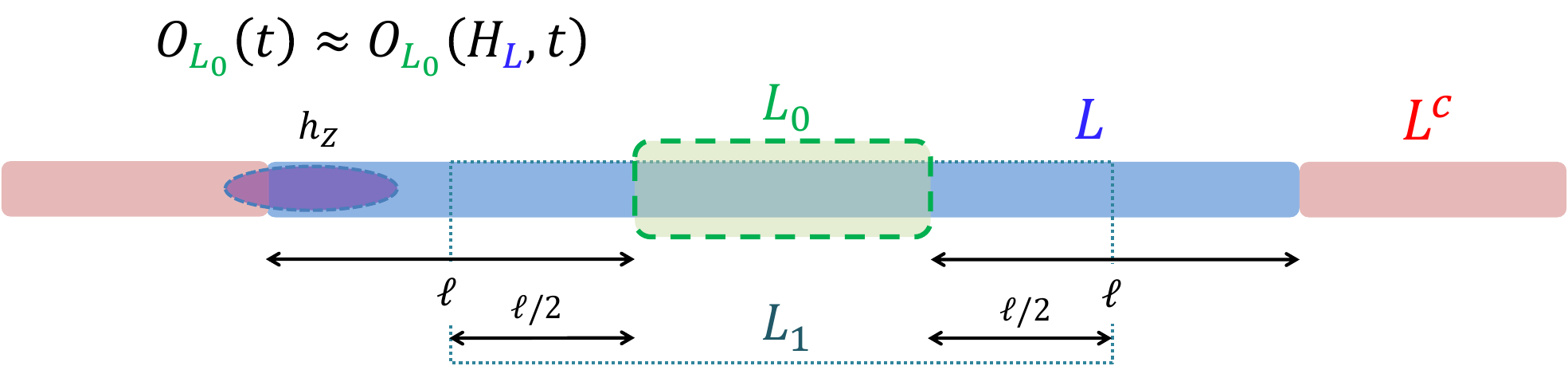}
\caption{The image illustrates an approximation of the time-evolved operator $O_{L_0}(t)$ by $O_{L_0}(H_L,t)$, where $L_0$ and $L^\co$ are separated by a distance $\ell$ (Corollary~\ref{corol_time_evo_approx}).
The approximation error is obtained by applying the Lieb--Robinson bound and Lemma~\ref{lem:Lieb-Robinson_start}, with the boundary interactions $h_Z$ such that $Z$ satisfies $Z \cap L \neq \emptyset$ and $Z\cap L^\co \neq \emptyset$. 
}
\label{fig:Lieb_Robinson}
\end{figure}

Furthermore, approximating $O_{L_0}(t)$ by the operator $O_{L_0}(H_L, t)$, using time evolution with respect to the subset Hamiltonian $H_L$ ($L_0 \subset L \subseteq \Lambda)$, can be useful in certain situations. As the distance between $L_0$ and $L^\co$ increases, the approximation error is expected to decrease. The following lemma confirms this (see Fig.~\ref{fig:Lieb_Robinson}):

\begin{lemma}[Supplementary Lemma~14 in Ref.~\cite{PhysRevLett.127.070403}] \label{lem:Lieb-Robinson_start}
Let $H$ be an arbitrary time-independent Hamiltonian of the form
\begin{align}
\label{lemma_Ham_tauH}
H = \sum_{Z\subset \Lambda} h_{Z}. 
\end{align}
Then, for any subset $L$ with $L \supseteq L_0$, the following inequality holds using the notation~\eqref{notatation_label_O_O',t}, i.e., $O(O',t) := e^{iO't} O e^{-iO't}$:
\begin{align}
\label{lem:ineq:Lieb-Robinson_start___0}
\| O_{L_0}(H, t) - O_{L_0}(H_{L,\tau},  t) \|  \le \sum_{Z: Z \in \mathcal{S}_{L}} \int_0^t \left\| [h_{Z,x}(H_L, x), O_{L_0}] \right\| dx ,
\end{align}
where $H_L$ represents the subset Hamiltonian on $L$, as introduced in \eqref{subset_Hamiltonian}, and $\mathcal{S}_{L}$ is defined as the collection of subsets $\{Z\}_{Z\subseteq \Lambda}$ that have a nonempty intersection with the boundary of $L$:
\begin{align}
\mathcal{S}_{L}:= \{ Z\subseteq \Lambda | Z \cap L \neq \emptyset, Z \cap L^\co \neq \emptyset\} .
\label{def_se_mathcak_S_L}
\end{align}
\end{lemma}

Using Lemma \ref{lem:Lieb-Robinson_start} and the Lieb--Robinson bound, we obtain the following corollary:
\begin{corol} \label{corol_time_evo_approx}
Let $\mathcal{F}(t,r)$ be the function introduced in \eqref{bound_comm_cor}. If the condition $\dist_{L^\co,L_0}\ge \ell$ is satisfied, the bound on $\| O_{L_0}(H, t) - O_{L_0}(H_L,t) \|$ in inequality~\eqref{lem:ineq:Lieb-Robinson_start___0} can be replaced with an explicit expression as follows:
\begin{align}
\label{lem:ineq:Lieb-Robinson_start}
\| O_{L_0}(H, t) - O_{L_0}(H_L,t) \| 
 \le |t| \cdot|L_0| \cdot \norm{O_{L_0} } \brr{ \frac{g \gamma^2 \ell^2}{2} \bar{J}(\ell/2) + \tilde{g}k \mathcal{F}(t,\ell/2)} ,
\end{align}
where $ \tilde{g} $ is defined as: 
\begin{align}
\label{def:tilde_g_0}
\tilde{g} := g\gamma^2 \bar{J}(1) .
\end{align}
\end{corol}

\textit{Proof of Corollary~\ref{corol_time_evo_approx}.}
We estimate the quantity  
\begin{align}
\sum_{Z: Z \in \mathcal{S}_{L}} \int_0^t \left\| [h_Z(H_L,x), O_{L_0}] \right\| dx .
\end{align}
from inequality~\eqref{lem:ineq:Lieb-Robinson_start___0}. We consider the decomposition 
\begin{align}
\label{deomp_h_Z_mathcal_S}
\sum_{Z: Z \in \mathcal{S}_{L}} h_Z = \sum_{Z: Z \in \mathcal{S}_{L}, \dist_{Z,L_0}\le \ell/2} h_Z + \sum_{Z: Z \in \mathcal{S}_{L}, \dist_{Z,L_0}>\ell/2} h_Z . 
\end{align}
It follows from the definition of $\mathcal{S}_{L}$ in Eq.~\eqref{def_se_mathcak_S_L} that 
\begin{align}
\label{S_L_upper_bound_H_Z}
 \sum_{Z: Z \in \mathcal{S}_{L}, \dist_{Z,L_0}\le \ell/2} \norm{h_Z}  \le 
  \sum_{Z: Z\cap L^\co\neq \emptyset, Z\cap L_1 \neq \emptyset } \norm{h_Z}  \le \bar{V}_{L^\co,L_1},
\end{align}
where we define the subset $L_1$ as $L_1:=\{i\in \Lambda| \dist_{i,L_0} \le \ell/2\}$ (see also Fig.~\ref{fig:Lieb_Robinson}). The upper bound $\bar{V}_{L^\co,L_1}$ was defined as in Eq.~\eqref{upper_bound_overline_H_X_l0}, and this $\bar{V}_{L^\co,L_1}$ is also bounded above by \eqref{definition_of_H_X_l0}. In the situation under discussion, upper bound \eqref{definition_of_H_X_l0} becomes
 \begin{align}
 \bar{V}_{L^\co,L_1}  \le g \gamma^2 (\ell/2)^2 \bar{J}(\ell/2) .   \label{definition_of_H_X_l0_ussing}
\end{align}
Therefore, we obtain 
\begin{align}
\label{lem:ineq:Lieb-Robinson_start_pro1}
 \sum_{Z: Z \in \mathcal{S}_{L}, \dist_{Z,L_0}\le \ell/2}  \int_0^t \left\| [h_Z(H_L,x), O_{L_0}] \right\| dx 
 \le 2 \cdot \frac{g \gamma^2 \ell^2}{4} |t|\cdot \norm{O_{L_0}} \bar{J}(\ell/2)  .
\end{align}

For the second term on the RHS of Eq.~\eqref{deomp_h_Z_mathcal_S}, using the Lieb--Robinson bound, we find that the following relations hold for $0\le x\le t$ and $\dist_{Z,L_0}>\ell/2$: 
\begin{align}
\left\| [h_Z(H_L,x), O_{L_0}] \right\|\le |Z| \cdot  |L_0| \cdot \norm{O_{L_0} } \cdot \norm{h_Z}\mathcal{F}(x,\dist_{Z,L_0}) \le  k|L_0| \cdot \norm{O_{L_0} } \cdot \norm{h_Z} \mathcal{F}(t,\ell/2), 
\end{align}
where we used $|Z|\le k$. 
Then, from the relations 
\begin{align}
 \sum_{Z: Z \in \mathcal{S}_{L}, \dist_{Z,L_0}> \ell/2} \norm{h_Z} 
 \le  \sum_{Z: Z \in \mathcal{S}_{L}} \norm{h_Z}  
 =  \sum_{Z: Z \cap L \neq \emptyset, Z \cap L^\co \neq \emptyset} \norm{h_Z}  
 = \bar{V}_{L,L^\co} \le   g \gamma^2 \bar{J}(1) =: \tilde{g} ,
\end{align}
we obtain 
\begin{align}
\label{lem:ineq:Lieb-Robinson_start_pro2}
\sum_{\substack{Z: Z \in \mathcal{S}_{L}, \dist_{Z,L_0}> \ell/2}} \int_0^t \left\| [h_Z(H_L,x), O_{L_0}] \right\| dx
\le \tilde{g}k|t| \cdot|L_0| \cdot \norm{O_{L_0} } \mathcal{F}(t,\ell/2) .
\end{align}
Combining inequalities~\eqref{lem:ineq:Lieb-Robinson_start_pro1} and \eqref{lem:ineq:Lieb-Robinson_start_pro2} proves the main inequality~\eqref{lem:ineq:Lieb-Robinson_start}. This completes the proof. $\square$

\subsection{Interaction-truncated Hamiltonian}  \label{sec:Interaction-truncated Hamiltonian}

\begin{figure}
\centering
\includegraphics[clip, scale=0.6]{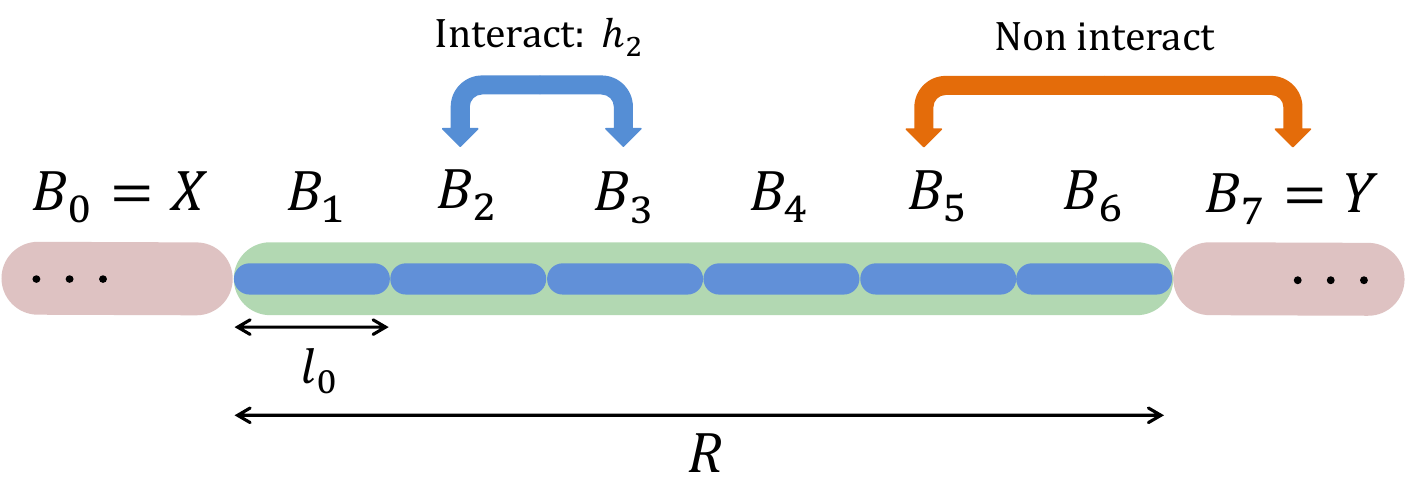}
\caption{We truncate the Hamiltonian's interactions by introducing subregions $\{B_s\}_{s=0}^{q+1}$ across the system. 
This results in partitionings of the system into $(q+2)$ blocks, with $q=6$, as depicted in the figure. 
Each block in the sequence of blocks $\{B_s\}_{s=1}^q$ has a length of $l_0$, while the terminal blocks $B_0$ and $B_{q+1}$ extend to the left and right extremities of the system, respectively. 
Following this, we truncate all interactions between these distinct blocks, which ensures the interaction length of at most $2l_0$. 
Then, the truncated Hamiltonian, denoted as $H_{\tc}$ in Eq.~\eqref{def:truncated_Hamiltonian}, remains substantially similar to the original Hamiltonian $H$, as established in Lemma~\ref{thm:locality_exp_effectiveHam}. 
We here consider the open boundary condition, but the same procedure can be applied under periodic boundary conditions (see Ref.~\cite[Supplementary Figure~6]{kuwahara2019area}).}
\label{Fig:Interaction_Truncate}
\end{figure}


We use the definition of the interaction-truncated Hamiltonian as provided in Ref.~\cite{kuwahara2019area} (see also Fig.~\ref{Fig:Interaction_Truncate}). 
We consider partitioning the total system into blocks: $B_0$, $\{B_s\}_{s=1}^q$, and $B_{q+1}$, satisfying $\bigcup_{s=0}^{q+1} B_s = \Lambda$. Here, $q$ is an even integer ($q \geq 2$), and we choose each $B_s$ (for $1 \leq s \leq q$) such that $|B_s| = l_0$. The subsets $X$ and $Y$ are defined in terms of these blocks as:
\begin{align}
    X=B_0, \quad Y = B_{q+1}. \label{Block_definition_LR}
\end{align}
We remark that, without loss of generality, for general regions $X$ and $Y$ one can include everything to the left of $X$ in $X$ and everything to the right of $Y$ in $Y$. In particular, $X$ and $Y$ can be huge, but the support of $O_X$ and $O_Y$ will not be included in the estimate.

We then truncate all interactions between non-adjacent blocks. 
After truncation, only interactions between adjacent blocks remain, leading to the truncated Hamiltonian:
\begin{align}
    H_\tc = \sum_{s=0}^{q+1} v_{s} + \sum_{s=0}^q h_s, \quad v_s:= H_{B_s} ,\quad h_s := V_{B_s,B_{s+1}}(B_s \sqcup B_{s+1}), 
     \label{def:truncated_Hamiltonian}
\end{align}
where we used the definition of $V_{L,L'}(\Lambda_0)$ as in Eq.~\eqref{Def:V_X_Y} with $X = B_s$, $Y = B_{s+1}$, and $\Lambda_0 = B_s \sqcup B_{s+1}$. $h_s$ denotes the interaction between blocks $B_s$ and $B_{s+1}$. 
Note that the relation $h_s \succeq 0$ holds because $h_Z \succeq 0$. 
From Lemma~\ref{lem_block_int} with $\dist_{B_s,B_{s+1}}=1$, we obtain the upper bound on $h_s$ as follows:
\begin{align}
\label{upper_bound_h_s/tilde_g}
\norm{h_s} = \norm{ V_{B_s,B_{s+1}}(B_s \sqcup B_{s+1})}\le g \gamma^2 \bar{J}(1) =: \tilde{g} . 
\end{align}
Applying Lemmas \ref{thm:locality_exp_effectiveHam} and \ref{lem:closeness_exp_op}, discussed below, we deduce the following inequality:
 \begin{align}
 \label{truncation:error_norm}
\norm{e^{\beta H} - e^{\beta H_\tc}}_1 
&\le 3 \beta \gamma^2 g q  l_0^2 \bar{J}(l_0) \tr \br{e^{\beta H}}  \notag \\
&= 3\gamma^2g\beta R l_0 \bar{J}(l_0) \tr \br{e^{\beta H}}  \for \beta \gamma^2 g q  l_0^2 \bar{J}(l_0) \le 1 ,
\end{align}
where we used $q=R/l_0$. 

\begin{lemma} \label{thm:locality_exp_effectiveHam}
The norm of the difference between $H$ and $H_\tc$ is bounded above:
 \begin{align}
\|\delta H_\tc\|\le  \gamma^2 g q  l_0^2 \bar{J}(l_0)   , \label{lemma:truncate__ineq}
\end{align}
where we define $\delta H_\tc := H-H_\tc$.
\end{lemma}
The bound in this lemma is a consequence of the upper bound in \eqref{definition_of_H_X_l0}. A proof of Lemma \ref{thm:locality_exp_effectiveHam} can be found in Ref.~\cite[Supplementary Lemma~3]{kuwahara2019area}.

{~}

\begin{lemma} \label{lem:closeness_exp_op}
For arbitrary Hermitian operators $A$ and $B$, the following inequality holds: 
 \begin{align}
 \label{lem:closeness_exp_op_main1}
\norm{e^{A+B}- e^{A}}_1 \le \|B\| e^{\norm{B}}\cdot \norm{e^{A}}_1 .
\end{align}
In particular, if $\|B\|\le 1$, this inequality reduces to
 \begin{align}
  \label{lem:closeness_exp_op_main2}
\norm{e^{A+B}- e^{A}}_1 \le 3\|B\| \cdot \norm{e^{A}}_1 .
\end{align}
\end{lemma}

{\bf Remark.} The inequality in Lemma \ref{lem:closeness_exp_op} is stronger than the trivial inequality 
 \begin{align}
\norm{e^{A+B}- e^{A}}_1= \norm{e^{A+B} e^{-A} e^{A}- e^{A}}_1
&\le  \norm{e^{A+B} e^{-A} -1}\cdot \norm{e^{A}}_1 \notag\\
&=   \norm{\mathcal{T} e^{\int_0^1 e^{\tau A}Be^{-\tau A} d\tau}  -1}\cdot \norm{e^{A}}_1 \notag\\
&\le e^{\norm{B} e^{2\norm{A}}}\norm{B} e^{2\norm{A}}\cdot \norm{e^{A}}_1 ,
\end{align}
which is meaningful only when $\norm{B}\ll e^{-2\norm{A}}$.

\textit{Proof of Lemma~\ref{lem:closeness_exp_op}.}
We consider the decomposition 
\begin{align}
e^{A+B} = \Phi_B e^{A} \Phi_B^\dagger ,
\end{align}
where $ \Phi_B$ is given by the belief propagation operator~\cite{kato2016quantum, PhysRevB.76.201102, PhysRevB.86.245116} (see also Lemma~\ref{BP:lemma_beta} below): 
\begin{align}
&\Phi_B = \mathcal{T} e^{\int_0^1 \phi_{B,\tau} d\tau} , \notag \\
&\phi_{B,\tau} =  \frac{1}{2}\int_{-\infty}^{\infty} f(t) B(A+\tau B, t)dt, \notag \\
&f(t):= \frac{1}{2\pi}\int_{-\infty}^{\infty} \tilde{f}(\omega) e^{-i\omega t}d\omega, \hspace{.15in} \tilde{f}(\omega)=\frac{{\rm tanh}(\omega/2)}{\omega/2} , 
\end{align}
and $\mathcal{T}$ is the time ordering operator. The function $f(t)$ is nonnegative, i.e., $f(t)\ge 0$, which becomes evident when rewritten in the explicit form Eq.~\eqref{explicit_form_of_f_beta_t} below. 
Hence, we have 
\begin{align}
\norm{\phi_{B,\tau}} \le  \frac{\norm{B}}{2}\int_{-\infty}^{\infty} f(t) dt = \frac{\norm{B}}{2}\tilde{f}(0)=  \frac{\norm{B}}{2} , 
\end{align}
which yields 
\begin{align}
\norm{ \Phi_B -\hat{1}} \le e^{\int_0^1\norm{\phi_{B,\tau} }d\tau} \int_0^1 \norm{\phi_{B,\tau}} d\tau \le  \frac{\norm{B}}{2} e^{\norm{B}/2} ,
\end{align}
where we applied \cite[Claim~25 therein]{PhysRevX.11.011047} to the first inequality, i.e.,
\begin{align}
\label{eq:PhysRevX.11.011047}
\norm{ \mathcal{T} e^{\int_0^1 A_\tau d\tau }-\mathcal{T} e^{\int_0^1 B_\tau d\tau }}\le 
e^{\int_0^1 \max \br{\norm{A_\tau} , \norm{B_\tau} }d\tau} \int_0^1 \norm{A_\tau-B_\tau} d\tau . 
\end{align}
We then obtain 
\begin{align}
\norm{e^{A+B} -e^A}_1 =  \norm{ \Phi_B e^{A} \Phi_B^\dagger - e^{A} \Phi_B^\dagger +e^{A} \Phi_B^\dagger  - e^A}_1
&\le  \norm{( \Phi_B-\hat{1}) e^{A} \Phi_B^\dagger}_1 +  \norm{e^{A} (\Phi_B^\dagger  -\hat{1})}_1  \notag \\
&\le \frac{\norm{B}}{2}  e^{\norm{B}/2} \br{1+ e^{\norm{B}/2}} \norm{e^{A}}_1  \notag \\
&\le \norm{B} e^{\norm{B}} \norm{e^{A}}_1  .
\end{align}
This completes the proof. $\square$

Truncating the Hamiltonian, an expression for the function $\mathcal{F}_0(r)$ \eqref{Lieb_robinson_func} in the Lieb--Robinson bound is given by:
\begin{corol}\label{LR_Lieb_Robinson:short}
For the interaction truncated Hamiltonian~\eqref{def:truncated_Hamiltonian}, the Lieb--Robinson bound is as follows:
 \begin{align}
 \label{int_truncate_Ham}
\norm{[O_{Z}(H_\tc, t), O_{Z'}]}\le \norm{O_{Z}} \cdot \norm{O_{Z'}} \cdot |Z| \cdot |Z'| e^{v|t|} \tilde{\mathcal{F}}(r) ,
\end{align}
for $Z,Z' \subset B_{1:q}$, where $r=\dist_{Z,Z'}$, and $\tilde{\mathcal{F}}(r)$ is given by 
\begin{align}
\label{Lieb_robinson_func_truncate}
\tilde{\mathcal{F}}(r)=C \min \br{ e^{-r/(2l_0)} , \bar{J}(r) } .
\end{align} 
\end{corol}
The fact that the interaction length is at most $2 l_0$ is used in the expression \eqref{Lieb_robinson_func_truncate}.

\subsection{Quantum belief propagation} \label{sec::Quantum belief propagation}

We adopt the interaction-truncated Hamiltonian introduced in the previous section. For notational simplicity, we denote $H_\tc$ as $H$, omitting the subscript. The Hamiltonian has the interaction length $2l_0$ and satisfies inequality~\eqref{int_truncate_Ham}. 

We introduce the notations $H_{\le s}$ and $H_{> s}$, defined as follows:
\begin{align}
H_{\le s} & := \sum_{s'\le s} v_{s'}+\sum_{s'<s} h_{s'}, \\\nonumber
H_{> s} & := \sum_{s'>s} \left(v_{s'}+h_{s'}\right).
\end{align}
The belief propagation operator is formulated as follows~\cite{kato2016quantum, PhysRevB.76.201102, PhysRevB.86.245116} :
\begin{lemma} \label{BP:lemma_beta}
Given the decomposition of the Hamiltonian as 
\begin{align}
H= H_{\le s} + H_{> s} + h_s , 
\end{align}
we obtain 
\begin{align}
e^{\beta H}=\Phi_s e^{\beta (H_{\le s} + H_{> s})}\Phi_s^\dagger, 
\end{align}
where the explicit expression for the belief propagation operator $\Phi_s$ is given as follows:
\begin{align}
\label{belief_prop_1}
\Phi_s & = \mathcal{T} e^{\int_0^1 \phi_{s,\tau}d\tau} ,\\\nonumber
\phi_{s,\tau} & = \frac{\beta}{2}\int_{-\infty}^{\infty} f_\beta(t) h_s(H_\tau, t)dt.
\end{align}
Here, $H_\tau:=H_{\le s} + H_{> s} + \tau h_s$.  
\end{lemma}
($\mathcal{T}$ and the function $f_\beta(t)$ in equations \eqref{belief_prop_1} were introduced in the proof of Lemma \ref{lem:closeness_exp_op} above.)
\vspace{.1in}

The function $f_\beta(t)$ was explicitly computed as follows \cite{Anshu_2021}:
\begin{align}
\label{explicit_form_of_f_beta_t}
f_\beta(t)=\frac{2}{\pi\beta}\log\left(\frac{e^{\pi|t|/\beta}+1}{e^{\pi|t|/\beta}-1}\right)\le \frac{2}{\pi\beta}\frac{2}{e^{\pi|t|/\beta}-1}.
\end{align}
Because $f_\beta(t) \ge 0$ and 
\begin{align}
\label{integrate_f_beta_t}
\int_{-\infty}^\infty| f_\beta(t)|dt= \tilde{f}(\omega=0) = 1,
\end{align}
we have 
\begin{align}
\label{belief_prop_upper_bound_phi_s}
\norm{\phi_{s,\tau}} \le  \frac{\beta}{2}\int_{-\infty}^{\infty} f_\beta(t) \norm{h_s(H_\tau, t)}dt \le  \frac{\beta\norm{h_s}}{2} \le \frac{\beta \tilde{g}}{2},
\end{align}
where we used inequality~\eqref{upper_bound_h_s/tilde_g}, i.e., $\norm{h_s}\le \tilde{g}$.
The above inequality immediately gives 
\begin{align}
\label{belief_prop_upper_bound_Phi_s_from_phi_s}
\norm{\Phi_s} & \le   e^{\int_0^1 \norm{\phi_{s,\tau}}d\tau}\le e^{\beta \tilde{g}/2}. 
\end{align}

\begin{figure}
\centering
\includegraphics[clip, scale=0.4]{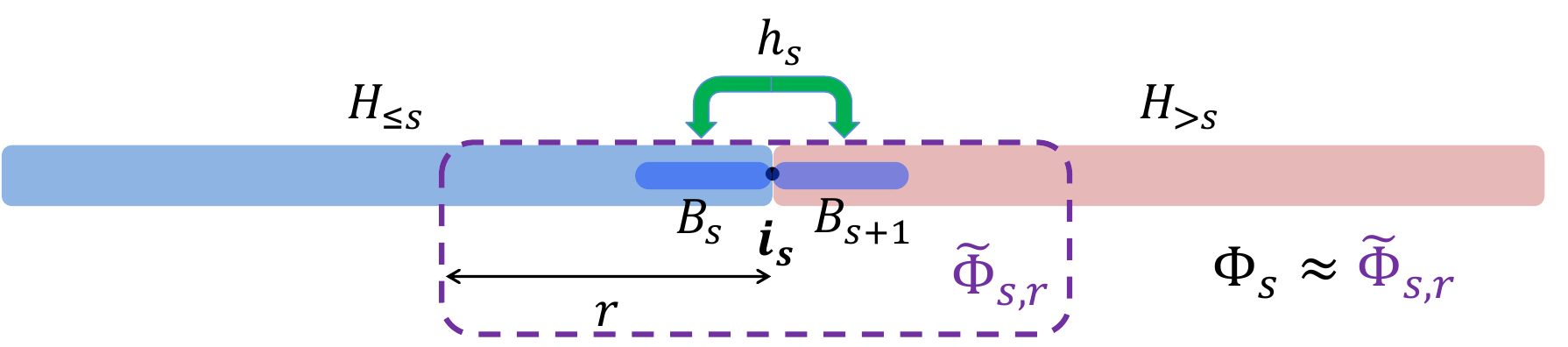}
\caption{The quantum belief propagation operator $\Phi_s$ characterizes the transformation from $e^{\beta H_{\le s}}  \otimes e^{\beta H_{> s}}$ to $e^{\beta H}$.
It is defined using time evolution with respect to $H_\tau:=H_{\le s} + H_{> s} + \tau h_s$. The Lieb--Robinson bound ensures the quasi-locality of $\Phi_s$ near the boundary, as shown in Proposition~\ref{lem:belief_error}. }
\label{fig:QBP}
\end{figure}

We introduce the notation $i_s$ to represent the boundary site that connects the blocks $B_s$ and $B_{s+1}$ (see Fig.~\ref{fig:QBP}).
We then consider the approximate belief propagation operator $\tilde{\Phi}_{s,\tilde{L}}$ onto the region $[i_s-r,i_s+r]$ (see Fig.~\ref{fig:QBP}), where $\tilde{\Phi}_{s,\tilde{L}}$ is defined by using the subset Hamiltonian $H_{\tilde{L}}$ with $\tilde{L}=[i_s-r,i_s+r]$ as follows:
\begin{align}
\label{def_tilde_Phi_phi}
\tilde{\Phi}_{s,\tilde{L}} := \mathcal{T} e^{\int_0^1 \tilde{\phi}_{s,\tau}d\tau} ,\quad  
\tilde{\phi}_{s,\tau} & = \frac{\beta}{2}\int_{-\infty}^{\infty} f_\beta(t) h_s\br{H_{\tau,\tilde{L}}, t}dt.
\end{align}
Here, $H_{\tau,\tilde{L}}:=H_{[i_s-r,i_s]} + H_{(i_s,i_s+r]} + \tau h_s$, supported on the region $\tilde{L}$.
We note that inequalities in forms identical to \eqref{belief_prop_upper_bound_phi_s} and \eqref{belief_prop_upper_bound_Phi_s_from_phi_s} hold for 
$\norm{\phi_{s,\tau}}$ and $\norm{\tilde{\Phi}_{s,\tilde{L}}}$, respectively. 
The error between the belief propagation operator $\Phi_s$ and the truncated operator $\tilde{\Phi}_{s,\tilde{L}} $, denoted by $\norm{\Phi_s - \tilde{\Phi}_{s,\tilde{L}}}$, has an upper bound that depends on $r$. We denote this upper bound by $\mathcal{G}_\beta(r)$:
\begin{align}
\norm{\Phi_s - \tilde{\Phi}_{s,\tilde{L}}} \le \mathcal{G}_\beta(r).
\end{align}
The form of $ \mathcal{G}_\beta(r)$ is estimated by combining the Lieb--Robinson bound and the form of the quantum belief propagation operator:
\begin{prop}  \label{lem:belief_error}
The norm of the approximation error between $\Phi_s$ and $\tilde{\Phi}_{s,\tilde{L}}$ has the following upper bound:
\begin{align}
\label{main:norm_Phi_s_tilde_Phi_s_r}
\norm{\Phi_s - \tilde{\Phi}_{s,\tilde{L}}} 
&\le e^{\beta \tilde{g}/2} \brrr{ \beta g \gamma^2 r^2 \br{1+  \frac{k\tilde{g} \beta}{2\pi^3}}  \bar{J}(r/3) 
+  \frac{k\tilde{g} \beta^2}{2\pi^3}\brr{9\tilde{g}  \sqrt{C\mathcal{F}_0(r/3)}  +72\tilde{g}  \br{\frac{\mathcal{F}_0(r/3)}{C}}^{\pi/(4v \beta)}} } \notag \\
&\le e^{\Theta(\beta)} \brr{r^2  \bar{J}(r/3)  +  \brr{ \mathcal{F}_0(r/3)}^{1/\Theta(\beta)}},
\end{align}
where we assume that $r$ satisfies $\mathcal{F}_0(r/3)\le 1$ and used the $\Theta$-notation as introduced in \eqref{Theta_notation_def}. 
In particular, if the truncated interaction length of $2l_0$ is smaller than $r/3$, i.e., 
\begin{align}
r > 6 l_0 \label{r>6l/_0_con},
\end{align}
the upper bound~\eqref{main:norm_Phi_s_tilde_Phi_s_r} reduces to the following form:
\begin{align}
\label{main:norm_Phi_s_tilde_Phi_s_r_trunc}
\norm{\Phi_s - \tilde{\Phi}_{s,\tilde{L}}} 
&= e^{\beta \tilde{g}/2} \frac{k\tilde{g} \beta^2}{2\pi^3}\brr{9\tilde{g}  \sqrt{\mathcal{F}_0(r/3)}  +72\tilde{g}  \brr{\frac{\mathcal{F}_0(r/3)}{C}}^{\pi/(4v \beta)}}   \notag \\
&\le e^{\Theta(\beta)} \brr{ \mathcal{F}_0(r/3)}^{1/\Theta(\beta)}  \notag \\
&= e^{\Theta(\beta)} \min \br{ e^{-r/[l_0\Theta(\beta)]} , \brr{\bar{J}(r/3)}^{1/\Theta(\beta)} } , 
\end{align}
where we used the expression \eqref{Lieb_robinson_func_truncate} for $\mathcal{F}_0(r/3)$. 
\end{prop}
Based on the above proposition, we define $\mathcal{G}_\beta(r)$ as 
\begin{align}
\label{def_mathcal_G}
\mathcal{G}_\beta(r)= e^{\Theta(\beta)} \min \br{ e^{-r/[l_0\Theta(\beta)]} , \brr{\bar{J}(r/3)}^{1/\Theta(\beta)} } ,
\end{align}  
which gives Therefore, $\norm{\Phi_s - \tilde{\Phi}_{s,\tilde{L}}} \le\mathcal{G}_\beta(r)$ for $r>6l_0$.


\subsubsection{Proof of Proposition~\ref{lem:belief_error}.}

By using inequality \eqref{eq:PhysRevX.11.011047}, we obtain 
\begin{align}
\label{norm_Phi_s_tilde_Phi_s_r}
\norm{\Phi_s - \tilde{\Phi}_{s,\tilde{L}}} \le e^{\beta \tilde{g}/2} \int_0^1 \norm{ \phi_{s,\tau} - \tilde{ \phi}_{s,\tau}}   d\tau ,
\end{align}
where we used $\max( \norm{ \phi_{s,\tau} } ,\norm{ \tilde{ \phi}_{s,\tau}} ) \le\beta \tilde{g}/2 $. 
Therefore, it suffices to deduce an upper bound on 
\begin{align}
\norm{ \phi_{s,\tau} - \tilde{ \phi}_{s,\tau} } .
 \end{align}

We approximate $h_s$ by restricting it to the region $[i_s-r/3,i_s+r/3]$. 
We then denote the approximated interaction by $h_{s,r/3}$. The approximation error is given by 
 \begin{align}
\norm{ h_s - h_{s,r/3}} \le  \norm{\bar{V}_{(-\infty, i_s-r/3), [i_s,\infty)}}+ \norm{\bar{V}_{(-\infty, i_s], (i_s+r/3,\infty)} } \le \frac{2g \gamma^2 r^2}{9} \bar{J}(r/3),
\end{align}
where $\overline{V}_{L,L'}$ was introduced as in Eq.~\eqref{upper_bound_overline_H_X_l0}, and the upper bound \eqref{definition_of_H_X_l0} holds for both $\overline{V}_{L,L'}$ and $\|V_{L,L'}(\Lambda_0)\|$. 
We introduce the following notations:
\begin{align}
\phi'_{s,\tau} =\frac{\beta}{2}\int_{-\infty}^{\infty} f_\beta(t)h_{s,r/3} (H_\tau,t) dt,\quad 
\tilde{\phi}'_{s,\tau}= \frac{\beta}{2}\int_{-\infty}^{\infty} f_\beta(t)h_{s,r/3} (H_{\tau,\tilde{L}},t) dt
\end{align}
with $\tilde{L}:=[i_s-r,i_s+r]$. Then, we have
\begin{align}
\label{norm_dif_phi_phi'}
\norm{\phi_{s,\tau} - \phi'_{s,\tau} } \le\frac{\beta}{2}\int_{-\infty}^{\infty} f_\beta(t) \norm{ h_s - h_{s,r/3}} dt 
&\le \frac{\beta g \gamma^2 r^2}{9} \bar{J}(r/3)\int_{-\infty}^{\infty} f_\beta(t) dt \notag \\
&= \frac{\beta g \gamma^2 r^2}{9} \bar{J}(r/3),
\end{align}
where we used Eq.~\eqref{integrate_f_beta_t} in the last equation.
An analogous inequality holds for $\norm{\tilde{\phi}_{s,\tau} - \tilde{\phi}'_{s,\tau} }$, where $\tilde{\phi}_{s,\tau}$ was defined as in Eq.~\eqref{def_tilde_Phi_phi}.

We then consider 
\begin{align}
\label{integral_hs}
\norm{\phi'_{s,\tau}- \tilde{\phi}'_{s,\tau} } \le\frac{\beta}{2}\int_{-\infty}^{\infty} f_\beta(t) \norm{h_{s,r/3}(H_\tau,t) - h_{s,r/3}(H_{\tau,\tilde{L}},t)} dt .
\end{align}
The norm $\norm{h_{s,r/3}(H_\tau,t) - h_{s,r/3}(H_{\tau,\tilde{L}},t)}$ on the RHS of \eqref{integral_hs} is bounded above as follows:
\begin{align}
\label{apprxo_H_tau_L_h_s_r/3}
\norm{ h_{s,r/3} (H_\tau,t)- h_{s,r/3} (H_{\tau,\tilde{L}},t)} \le 
\sum_{\substack{Z: Z\subseteq [i_s-r/3,i_s+r/3] \\
Z\cap (-\infty,s] \neq \emptyset, Z\cap (i_s,\infty) \neq \emptyset}} \norm{ h_Z(H_\tau,t)-  h_Z(H_{\tau,\tilde{L}},t)}.
\end{align}
Since $\dist_{Z,L^\co}\ge 2r/3$, by using Corollary~\ref{corol_time_evo_approx}, we obtain 
\begin{align}
\norm{ h_Z(H_\tau,t)-  h_Z(H_{\tau,\tilde{L}},t)} \le |t| k \norm{h_Z} \brr{ \frac{g \gamma^2 r^2}{9}  \bar{J}(r/3) + \tilde{g}\mathcal{F}(t,r/3)} ,
\end{align}
which simplifies inequality~\eqref{apprxo_H_tau_L_h_s_r/3} to 
\begin{align}
\label{apprxo_H_tau_L_h_s_r/3_fin}
\norm{ h_{s,r/3} (H_\tau,t)- h_{s,r/3} (H_{\tau,\tilde{L}},t)} \le 
|t| k \tilde{g} \brr{ \frac{g \gamma^2 r^2}{9}  \bar{J}(r/3) + \tilde{g}\mathcal{F}(t,r/3)}  ,
\end{align}
where we used inequality~\eqref{upper_bound_h_s/tilde_g}. 
Thus, we have
\begin{align}
\label{upper_bound_h_s/tilde_g__re}
\norm{\phi'_{s,\tau}- \tilde{\phi}'_{s,\tau} } \le\frac{\beta}{2}\int_{-\infty}^{\infty} f_\beta(t)|t| k \tilde{g} \brr{ \frac{g \gamma^2 r^2}{9}\bar{J}(r/3) + \tilde{g}\mathcal{F}(t,r/3)}  dt .
\end{align}

Finally, we compute the integral on the RHS of Eq.~\eqref{upper_bound_h_s/tilde_g__re}. 
From definition~\eqref{explicit_form_of_f_beta_t}, we have 
\begin{align}
\int_{-\infty}^\infty  |t| f_\beta(t)dt = \frac{4\beta}{\pi^3} \int_0^\infty x \log \br{\frac{e^x+1}{e^x-1}} dx =  \frac{7\zeta(3)\beta}{\pi^3} ,
\end{align}
and hence,
\begin{align}
\label{upper_bound_h_s/tilde_g__re_re}
\norm{\phi'_{s,\tau}- \tilde{\phi}'_{s,\tau} }
 \le   \frac{k \tilde{g} \beta^2}{2\pi^3} \cdot \frac{7\zeta(3)g \gamma^2 r^2}{9}\bar{J}(r/3) +  \frac{\beta}{2}\int_{-\infty}^{\infty} f_\beta(t)|t| k \tilde{g} \brr{ \tilde{g}\mathcal{F}(t,r/3)}  dt ,
\end{align}
where $\zeta(x)$ represents the Riemann zeta function. 
We next consider 
\begin{align}
\label{integrate_f_beta_t_r_3}
\int_{-\infty}^{\infty} |t| f_\beta(t) \mathcal{F}(t,r/3) dt 
&= 2\int_{0}^{\infty} t f_\beta(t) \mathcal{F}(t,r/3) dt   \notag \\
&=2\int_{0}^{t_0} t f_\beta(t) \mathcal{F}(t,r/3) dt +2 \int_{t_0}^{\infty} t f_\beta(t) \mathcal{F}(t,r/3) dt  .
\end{align}
In the above decomposition, $t_0$ is chosen so that the following inequality holds for $t\le t_0$:
\begin{align}
e^{vt} \mathcal{F}_0(r/3) \le \sqrt{C\mathcal{F}_0(r/3)} .  
\end{align}
We can find such $t_0\ge 0$ since the relation $\mathcal{F}_0(r/3)\le C$ holds, as can be seen from the definition Eq.~\eqref{Lieb_robinson_func}. Note that $\mathcal{F}(t,r/3)\le e^{vt} \mathcal{F}_0(r/3)$ as in Eq.~\eqref{def:mathcal_F_(r)}.  
Solving this inequality for $t$, we deduce that $t_0$ takes the following value: 
\begin{align}
t_0= \frac{-1}{2v} \log\left(\mathcal{F}_0(r/3)/C\right).
\end{align}
We then obtain 
\begin{align}
\label{integrate_f_beta_t_r_3_1}
2\int_{0}^{t_0} t f_\beta(t) \mathcal{F}(t,r/3) dt \le 2 \sqrt{C\mathcal{F}_0(r/3)} \int_{0}^{\infty} t f_\beta(t) dt =  \frac{7 \zeta(3)\beta}{\pi^3} \sqrt{C\mathcal{F}_0(r/3)} .
\end{align}

Due to what was previously mentioned below Lemma \ref{lemma_3}, for the region $t\ge t_0$, we have $\mathcal{F}(t,r/3)\le 2$, 
and hence, we have 
\begin{align}
2 \int_{t_0}^{\infty} t f_\beta(t) \mathcal{F}(t,r/3) dt \le 4 \int_{t_0}^{\infty} t f_\beta(t) dt .
\end{align}
For this quantity, we find that the following relations hold:
\begin{align}
\label{integrate_f_beta_t_r_3_2}
4 \int_{t_0}^{\infty} t f_\beta(t) dt
&\le 4  e^{-\pi t_0/(2\beta) }\int_{t_0}^{\infty} t e^{\pi t/(2\beta) } f_\beta(t)dt  \le \frac{72\beta}{\pi^3} \brr{\frac{\mathcal{F}_0(r/3)}{C}}^{\pi/(4v \beta)},
\end{align}
where we used 
\begin{align}
\int_{0}^\infty  te^{\pi t/(2\beta) } f_\beta(t)dt = \frac{2\beta}{\pi^3} \int_0^\infty x e^{x/2}\log \br{\frac{e^x+1}{e^x-1}} dx 
\le  \frac{18\beta}{\pi^3} .
\end{align}

By applying inequalities~\eqref{integrate_f_beta_t_r_3_1} and \eqref{integrate_f_beta_t_r_3_2} to Eq.~\eqref{integrate_f_beta_t_r_3}, we obtain 
\begin{align}
\label{integrate_f_beta_t_r_3_fin}
\int_{-\infty}^{\infty} |t| f_\beta(t) \mathcal{F}_0(r/3) dt 
&\le  \frac{\beta}{\pi^3} \br{7\zeta(3) \sqrt{C\mathcal{F}_0(r/3)}  +72 \brr{\frac{\mathcal{F}_0(r/3)}{C}}^{\pi/(4v \beta)} } ,
\end{align}
which reduces inequality~\eqref{upper_bound_h_s/tilde_g__re_re} to
\begin{align}
\norm{\phi'_{s,\tau}- \tilde{\phi}'_{s,\tau} } 
&\le \frac{k\tilde{g} \beta^2}{2\pi^3}\brr{ \frac{7\zeta(3) g \gamma^2 r^2}{9}\bar{J}(r/3) + \tilde{g} \br{7\zeta(3) \sqrt{C\mathcal{F}_0(r/3)}  +72 \brr{\frac{\mathcal{F}_0(r/3)}{C}}^{\pi/(4v \beta)}}} \notag \\
&\le \frac{k\tilde{g} \beta^2}{2\pi^3}\brr{g \gamma^2 r^2 \bar{J}(r/3) + 9\tilde{g}  \sqrt{C\mathcal{F}_0(r/3)}  +72\tilde{g}  \brr{\frac{\mathcal{F}_0(r/3)}{C}}^{\pi/(4v \beta)}},
\end{align}
where we use $7\zeta(3)\approx 8.4144<9$.
Combining this bound with~\eqref{norm_dif_phi_phi'}, we finally arrive at the inequality 
\begin{align}
\norm{\phi_{s,\tau} - \tilde{\phi}_{s,\tau} } 
&\le
\norm{\phi_{s,\tau} - \phi'_{s,\tau} } + \norm{ \tilde{\phi}_{s,\tau} -  \tilde{\phi}'_{s,\tau} }  + \norm{\phi'_{s,\tau}- \tilde{\phi}'_{s,\tau} }  \notag \\
&\le  \frac{2\beta g \gamma^2 r^2}{9} \bar{J}(r/3)+ \frac{k\tilde{g} \beta^2}{2\pi^3}\brrr{g \gamma^2 r^2 \bar{J}(r/3) + 9\tilde{g}  \sqrt{\mathcal{F}_0(r/3)}  +72\tilde{g}  \brr{\frac{\mathcal{F}_0(r/3)}{C}}^{\pi/(4v \beta)}} . 
\end{align}
Now the main inequality~\eqref{main:norm_Phi_s_tilde_Phi_s_r} follows from inequality~\eqref{norm_Phi_s_tilde_Phi_s_r}.

The inequality \eqref{norm_dif_phi_phi'} implies the equation of $\bar{J}(r/3)=0$ under condition \eqref{r>6l/_0_con} and the interaction length $2l_0$, reducing the above inequality to the form of \eqref{main:norm_Phi_s_tilde_Phi_s_r_trunc}. 
This completes the proof. 
$\square$

\section{Cluster expansion technique} \label{sec:Cluster expansion technique}

\begin{figure}
\centering
\includegraphics[clip, scale=0.4]{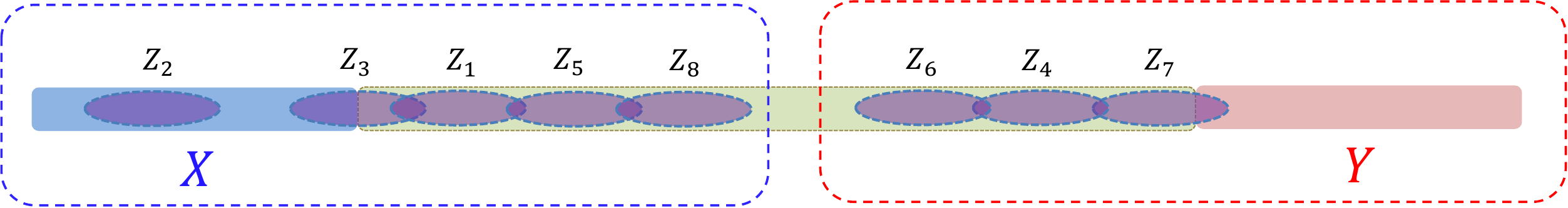}
\caption{An image displaying the decomposition of the collection of sets $\{X, Z_1, Z_2, \ldots, Z_8, Y\}$ into two disjoint sub-collections, $\{X, Z_1, Z_2, Z_3, Z_5, Z_8 \}$ and $\{Z_4, Z_6,Z_7, R \}$, is shown. 
When the sets $Z_1$, $Z_2$, $\ldots$, $Z_8$ cannot connect the distant regions $L$ and $R$, the trace vanishes.}
\label{fig:connection_subset}
\end{figure}

In proving the clustering theorem, we utilize the cluster expansion technique. 
We first rewrite an arbitrary bipartite correlation function $\Cor_{\rho_\beta}(O_X,O_Y)$ by following~\cite{PhysRevX.4.031019}.
For this purpose, we adopt a copy of the original Hilbert space and define the notations $O^{(+)}$, $O^{(0)}$ and $O^{(1)}$ for an arbitrary operator $O$ as follows: 
 \begin{align}
O^{(+)} = O \otimes \hat{1} +  \hat{1} \otimes O , \quad O^{(0)} = O \otimes \hat{1} , \quad  O^{(1)} =  O \otimes \hat{1} -  \hat{1} \otimes O.
\end{align}
Using the above notation, we obtain 
\begin{align}
\label{eq:cluster_form}
\Cor_{\rho_\beta}(O_X,O_Y) = \tr \br{ \rho_\beta\otimes \rho_\beta O_X^{(0)} O_Y^{(1)}  } =\frac{1}{Z_\beta^{(+)}}\tr \br{ e^{\beta H^{(+)}} O_X^{(0)} O_Y^{(1)}  } ,
\end{align}
where $Z_\beta^{(+)}=\tr \br{ e^{\beta H^{(+)}}}=Z_\beta^2 $.
One can easily verify that Eq.~\eqref{eq:cluster_form} holds. 
Then, the following lemma applies:
\begin{lemma} \label{lem:cluster_exp}
\label{claim}
Given subsets $X$, $Y$, and $Z_i$ for $i=1, \ldots, m$, we assume that $X$ and $Y$ are disjoint, i.e., $X\cap Y=\emptyset$, and that the collection of sets $\{X, Z_1,Z_2,\ldots, Z_m,Y\}$ admits a decomposition into two sub-collections such that the union of the sets in one sub-collection and the union of the sets in the other sub-collection are disjoint (see Fig.~\ref{fig:connection_subset}); that is, there exists a decomposition $ \{1,2,\ldots,m\}= \{i_1,i_2, \ldots, i_s\} \sqcup \{i_{s+1},i_{s+2}, \ldots, i_m\}$ such that 
\begin{align} 
\label{decomp_sets}
\br{X\cup Z_{i_1} \cup Z_{i_2} \cup \cdots \cup Z_{i_s}} \cap \br{Y\cup Z_{i_{s+1}} \cup Z_{i_{s+2}} \cup \cdots \cup Z_{i_m}} = \emptyset  . 
\end{align}
Then, for the set of operators $\{O_{Z_j}\}_{j=1}^m$, the following relation holds: 
\begin{align}
\tr \br{ O_{Z_1}^{(+)} O_{Z_2}^{(+)} \cdots O_{Z_m}^{(+)}   O_X^{(0)} O_Y^{(1)}  } =0,
\end{align}
\end{lemma}

\textit{Proof of Lemma~\ref{lem:cluster_exp}.}
By assumption, the collection of sets $\{X, Z_1, \ldots, Z_m, Y\}$ decomposes into two sub-collections of sets:
\begin{align}
\mathcal{C}_1 & =\{X, Z_{i_1}, \ldots, Z_{i_s}\}, \\\nonumber
\mathcal{C}_2 & =\{Y, Z_{i_{s+1}}, \ldots, Z_{i_m}\},
\end{align}
such that the union of the sets in the sub-collection $\mathcal{C}_1$ and the union of the sets in the sub-collection $\mathcal{C}_2$ are disjoint, as in relation \eqref{decomp_sets}. This means that the trace
\begin{align}
\label{prdt_tr}
{\rm tr} \left(O^{(+)}_{Z_1} O^{(+)}_{Z_2}\ldots O^{(+)}_{Z_m} O^{(0)}_X O^{(1)}_Y\right)
\end{align}
can be rewritten as the product of traces as follows:
\begin{align}
{\rm tr} \left(O^{(+)}_{Z_1} O^{(+)}_{Z_2}\ldots O^{(+)}_{Z_m} O^{(0)}_X O^{(1)}_Y\right)= {\rm tr} \left(\prod_{j=1}^s O^{(+)}_{Z_{i_j}}\cdot O^{(0)}_X \right)\, {\rm tr} \left(O^{(1)}_Y \prod_{k=s+1}^m O^{(+)}_{Z_{i_k}}\right).
\end{align}

However, the operators $ O^{(+)}_{Z_{i_k}}=O_{Z_{i_k}}\otimes \hat{1}+\hat{1}\otimes O_{Z_{i_k}}$ are symmetric, while the operator $O^{(1)}_Y=O_Y\otimes \hat{1}-\hat{1}\otimes O_Y$ is antisymmetric under the exchange between the two Hilbert spaces. Owing to these properties of the operators, we have the following equalities:
\begin{align}
{\rm tr} \left(O^{(1)}_Y \prod_{k=s+1}^m O^{(+)}_{Z_{i_k}} \right)= {\rm tr} \left(O_Y\otimes\hat{1}\cdot \prod_{k=s+1}^m O^{(+)}_{Z_{i_k}}\right)- {\rm tr} \left(\hat{1}\otimes O_Y\cdot\prod_{k=s+1}^m O^{(+)}_{Z_{i_k}}\right)=0.
\end{align}
This shows that the trace \eqref{prdt_tr} vanishes under the assumption. 
This completes the proof.  $\square$

{~}

By using Lemma \ref{claim}, the function $\Cor_{\rho_\beta}(O_X,O_Y)$ can be expanded as 
\begin{align}
\Cor_{\rho_\beta}(O_X,O_Y) =\frac{1}{Z_\beta^{(+)}}\sum_{m=0}^\infty \sum_{Z_1,Z_2,\ldots,Z_m: {\rm connected}} \frac{1}{m!} \tr \br{h_{Z_1}^{(+)} h_{Z_2}^{(+)} \cdots h_{Z_m}^{(+)}O_X^{(0)} O_Y^{(1)}  },
\end{align}
where the summation $\sum_{Z_1, Z_2,\ldots, Z_m: {\rm connected}}$ is taken over all possible collections $\{Z_1,Z_2,\ldots, Z_m\}$ such that the collection $\{X, Z_1,Z_2,\ldots, Z_m,Y\}$ connects $L$ and $R$. 
We then define the operator $\rho_{\cl}$ as follows:
\begin{align}
\label{correlation_expansion__2}
\rho_{\cl}:=\frac{1}{Z_\beta^{(+)}} \sum_{m=0}^\infty \sum_{Z_1,Z_2,\ldots,Z_m: {\rm connected}} \frac{\beta^m}{m!} h_{Z_1}^{(+)} h_{Z_2}^{(+)} \cdots h_{Z_m}^{(+)}. 
\end{align}
Using the operator $\rho_{\cl}$, the bipartite correlation ${\rm Cor}_{\rho_\beta}(O_X, O_Y)$ has the following upper bound: 
\begin{align}
\label{correlation_up_cluster}
\Cor_{\rho_\beta}(O_X,O_Y) \le \norm{O_X} \cdot \norm{O_Y} \cdot \norm{\rho_{\cl}}_1 .
\end{align}

A standard counting argument (e.g., in Refs.~\cite{Kotecky1986,PRXQuantum.5.010305}) gives an upper bound on the summand in Eq.~\eqref{correlation_expansion__2} as 
\begin{align}
\label{upper_bound_beta_m}
\frac{1}{Z_\beta^{(+)}} \sum_{Z_1,Z_2,\ldots,Z_m: {\rm connected}} \frac{1}{m!} \norm{h_{Z_1}^{(+)} h_{Z_2}^{(+)} \cdots h_{Z_m}^{(+)} }
\lesssim  \br{\beta/\beta_c}^m ,
\end{align}
where $\beta_c$ is a constant that depends on the system detail.
The upper bound on the RHS of \eqref{upper_bound_beta_m} summed over $m$ converges only when the value of $\beta$ is less than the threshold value, $\beta_c$, i.e., only when $\beta< \beta_c$. Therefore, this upper bound ensures that the summation in $\rho_{\cl}$ converges to a finite value only when $\beta < \beta_c$. 
This limitation presents the main challenge in proving the clustering theorem in a 1D system at arbitrary temperatures. 
Therefore, this argument does not apply to inverse temperatures above the threshold value $\beta_c$. Given this context, we take a different approach to finding a resolution.  

In the following, we focus on the cluster expansion in 1D finite-range interacting systems as in Eq.~\eqref{def:truncated_Hamiltonian}. 
We define the operator $\Psi_{X,Y}$ as 
\begin{align}
\label{def: Psi_X_Y}
\Psi_{X,Y}:= O_X^{(0)}  O_Y^{(1)},\quad \norm{\Psi_{X,Y}} \le 2 ,
\end{align}
where, using $\norm{O_X^{(0)}}\le \norm{O_X}=1$ and $\norm{O_Y^{(1)}}=\norm{O_Y\otimes \hat{1} - \hat{1} \otimes O_Y} \le 2\norm{O_Y}=2$, we immediately find that $\norm{\Psi_{X,Y}} \le 2$. 

In the following discussions, for notational simplicity, we denote $H^{(+)}$ as $H$, and $Z_\beta^{(+)}$ as $Z_\beta$, omitting the superscripts. As we mentioned previously, we have denoted $H_\tc$ as $H$.
This notational convention only alters the norm of the local interactions by a factor of two. 
Since we have $\norm{h^{(+)}_s} \le 2 \norm{h_s}  \le 2 \tilde{g}$, we need to use the upper bound
\begin{align}
\label{new_h_s_bound}
\norm{h_s} \le 2 \tilde{g} 
\end{align}
when $h^{(+)}_s$ is abbreviated as $h_s$.  

Then, if one of the interaction terms $\{h_s\}_{s=0}^q$ has a zeroth-order term in the Taylor expansion of $e^{\beta H}$, the cluster is disconnected [$h_s$ was introduced in \eqref{def:truncated_Hamiltonian}].
To describe this mathematically, we define the operator  
\begin{align}
G_s :=  e^{\beta H}-e^{\beta (H-h_s)}= -\int_0^1  \frac{d}{d\lambda_s}e^{\beta H - \beta \lambda_s h_s} d\lambda_s .
\end{align}
By construction, the operator $G_s$ removes the zeroth-order term of $h_s$ from the Taylor expansion of $e^{\beta H}$. 
This generalizes to situations involving multiple parameters. In the case of multiple parameters, we define 
\begin{align}
\label{op_G_j_1___s_m}
G_{s_1,s_2,\ldots,s_m} 
&:=(-1)^m\int_0^1\int_0^1\cdots \int_0^1d\lambda_{s_1}d\lambda_{s_2}\cdots d\lambda_{s_m}   \frac{d}{d\lambda_{s_1} } \frac{d}{d\lambda_{s_2} } \cdots \frac{d}{d\lambda_{s_m} }  
 e^{\beta H -\beta \sum_{j=1}^m \lambda_{s} h_{s_j}} \notag \\
&=\mathcal{D}_{h_{s_1},\ldots,h_{s_m}}e^{\beta H} .
 \end{align}
The super-operator $\mathcal{D}_{h_{s_1},\ldots,h_{s_m}}$ removes all zeroth-order terms of $h_{s_j}$ ($j=1, \ldots, m$) from the Taylor expansion of $e^{\beta H}$. 
Using Lemma~\ref{lem:cluster_exp}, we obtain 
\begin{align}
\label{correlation_up_cluster_1D_90}
\tr\brr{\Psi_{X,Y}  \br{\rho_\beta- G_{s_1,s_2,\ldots,s_m}}} =0
\end{align}
for an arbitrary set $\{s_1,s_2,\ldots,s_m\}$. From this, we deduce the following inequalities:
\begin{align}
\label{correlation_up_cluster_1D}
\Cor_{\rho_\beta}(O_X,O_Y) \le\frac{1}{Z_\beta} \abs{\tr\br{\Psi_{X,Y} G_{s_1,s_2,\ldots,s_m}}} \le \frac{2}{Z_\beta} \norm{ G_{s_1,s_2,\ldots,s_m}}_1 .
\end{align}
Therefore, upper-bounding the norm $\norm{G_{s_1,s_2,\ldots,s_m} }_1$ leads to the clustering theorem.

\subsection{1D cases: commuting Hamiltonian}\label{1D cases: commuting Hamiltonian}

We demonstrate the commutative situation here, i.e., $[h_s,h_{s'}]=0$ $\forall s,s' \in [0,q]$ in Eq.~\eqref{def:truncated_Hamiltonian}. The aforementioned restriction on the values of $\beta$ does not apply in this simple scenario. Comparing the commutative and noncommutative situations allows for a clear identification of the challenges that arise in proving the clustering theorem. The challenges for noncommutative situations will be identified afterwards.
%
%

In the commutative case, the operation $\mathcal{D}_{h_{s_1},\ldots,h_{s_m}}$ in Eq.~\eqref{op_G_j_1___s_m} 
simply replaces $e^{\beta h_{s_1}}$ with $e^{\beta h_{s_1}}-\hat{1}$ in the Gibbs state 
\begin{align}
e^{\beta H} = \prod_{s=0}^{q+1} e^{\beta v_s}  \prod_{s=0}^{q} e^{\beta h_s}  ,
\end{align}
where we used the notations defined in Eq.~\eqref{def:truncated_Hamiltonian}. 
The trace norm $\norm{G_{0,1,\ldots,q} }_1$ is simply the trace of $G_{0,1,\ldots,q} $: 
\begin{align}
\label{trace_norm_G_replce}
\norm{G_{0,1,\ldots,q} }_1 = \tr \br{G_{0,1,\ldots,q} } =\tr \brr{\prod_{s=0}^{q+1} e^{\beta v_s}  \prod_{s=0}^{q} \br{e^{\beta h_s} -\hat{1}} }  . 
\end{align}
Note that the positivity of the operator $\br{e^{\beta h_s} -\hat{1}}$ is ensured by $h_s \succeq 0$. 

To obtain an upper bound on this quantity, as an initial step, we observe the following inequality:
\begin{align}
\tr \brr{ e^{\beta H- \beta h_0} \br{e^{\beta h_0} -\hat{1}}} 
= \tr \br{ e^{\beta H} } \cdot \frac{\tr \brr{ e^{\beta H- \beta h_0} \br{e^{\beta h_0} -\hat{1}}} }{\tr \left(e^{\beta H- \beta h_0} e^{\beta h_0}\right)} 
\le \tr \left(e^{\beta H}\right) \cdot \frac{e^{\beta ||h_0||} -1}{e^{\beta ||h_0||}},
\end{align}
where the last inequality holds owing to the fact that for any quantum state $\rho$, the inequality
\begin{align}
\frac{\tr \brr{ \rho (e^{\beta h_0} -\hat{1})} }{\tr \left(\rho e^{\beta h_0}\right)} \le \frac{e^{\beta ||h_0||} -1}{e^{\beta ||h_0||}}
 \end{align}
holds.
In the second step, the same procedure yields the inequalities
\begin{align}
\label{cal_up_h_1_h_2_zeroth}
&\tr \brr{ e^{\beta (H-h_0-h_1)} \br{e^{\beta h_0} -\hat{1}} \br{ e^{\beta h_1} -\hat{1}} } \notag \\
&= \tr \brr{ e^{\beta (H-h_0)}\br{e^{\beta h_0} -\hat{1}}}  
\cdot \frac{\tr \brr{ e^{\beta (H-h_0-h_1)}\br{e^{\beta h_0} -\hat{1}}\br{ e^{\beta h_1} -\hat{1}}} }{\tr \brr{ e^{\beta (H-h_0-h_1)}\br{e^{\beta h_0} -\hat{1}}e^{\beta h_1}} } \notag \\
&\le\tr \brr{ e^{\beta (H-h_0)} \br{e^{\beta h_0} -\hat{1}}  }\cdot \frac{e^{\beta ||h_1||} -1}{e^{\beta ||h_1||}} \notag \\
&\le \tr \left(e^{\beta H}\right) \cdot  \frac{e^{\beta ||h_0||} -1}{e^{\beta ||h_0||}}  \cdot  \frac{e^{\beta ||h_1||} -1}{e^{\beta ||h_1||}} ,
\end{align}
where in the first inequality, we used the fact that $e^{\beta (H-h_0-h_1)} \br{e^{\beta h_0} -\hat{1}}$ is still positive semidefinite since $e^{\beta h_0} -\hat{1} \succeq 0$. 
In an analogous manner, after iterating this method, one deduces the following relation:
\begin{align}
\label{ineq_general_d}
\tr \brr{\prod_{s=0}^{q+1} e^{\beta v_s}  \prod_{s=0}^{q} \br{e^{\beta h_s} -\hat{1}} } \le \tr \left(e^{\beta H}\right) \cdot \prod_{s=0}^{q} \frac{e^{\beta ||h_s||} -1}{e^{\beta ||h_s||}}.
\end{align}

Using inequality~\eqref{correlation_up_cluster_1D}, we obtain the following upper bound on the correlation function:  
\begin{align}
\label{correlation_up_cluster_1D_commut}
\Cor_{\rho_\beta}(O_X,O_Y) \le 2 \prod_{s=0}^{q} \frac{e^{\beta ||h_s||} -1}{e^{\beta ||h_s||}} .
\end{align}
We obtained an upper bound on $||h_s||$ in~\eqref{new_h_s_bound}: 
\begin{align}
||h_s||\le 2\tilde{g},
\end{align}
therefore, we have 
\begin{align}
\Cor_{\rho_\beta}(O_X,O_Y) 
\le 2 \prod_{s=0}^{q} \frac{e^{2\beta \tilde{g}} -1}{e^{2\beta \tilde{g}}}
&= 2 \br{1- e^{-2\beta \tilde{g}} }^{q+1} \le 2 e^{-R/ \br{l_0 e^{2\beta \tilde{g}}}} ,
\label{bnd_rho_g}
\end{align}
where we used $q=R/l_0$ (this relation was introduced in Sec.~\ref{sec:Interaction-truncated Hamiltonian}) and the fact that for $x>0$, the inequality  
\begin{align}
1-\frac{1}{x} \le e^{-\frac{1}{x}} 
\end{align} 
holds. We thus conclude that this proves the clustering theorem for the commutative case with the correlation length $l_0 e^{2\beta \tilde{g}}$.
When the interaction length $l_0$ is a constant independent of $\beta$, the correlation length is given by $e^{{\rm const.} \times \beta}$.

When extending this discussion to the non-commuting Hamiltonian, we need to address the following problems.
To clarify the discussion, we explicitly express the operator $G_{s_1,s_2}$ in Eq.~\eqref{op_G_j_1___s_m} as follows: 
\begin{align}
G_{s_1,s_2}= e^{\beta H} - e^{\beta (H-h_{s_1})}- e^{\beta (H-h_{s_2})} + e^{\beta (H-h_{s_1}-h_{s_2})}.
\end{align}
The first difficulty arises because
\begin{align}
\label{first_diff_non_comm}
e^{\beta H} - e^{\beta (H-h_{s_1})}- e^{\beta (H-h_{s_2})} + e^{\beta (H-h_{s_1}-h_{s_2})} \neq 
e^{\beta (H-h_{s_1}-h_{s_2})} \br{1- e^{\beta h_{s_1}} } \br{1- e^{\beta h_{s_2}} }, 
\end{align}
and LHS is not equal to RHS since the operators $h_{s_1}$ and $h_{s_2}$ do not commute. 
This fact prevents us from using the prescription as in~\eqref{cal_up_h_1_h_2_zeroth}.

The second difficulty arises from the fact that, in general,  
\begin{align}
| G_{s_1,s_2} | \neq G_{s_1,s_2}. 
\end{align}
This inequality stems from the positivity of $h_{s_1}$ and $h_{s_2}$. 
To explain this point in more detail, the relation
\begin{align}
e^{\beta H}-e^{\beta H-\beta (h_{s_1}+\zeta) }\succeq 0 \label{beta_H_non-positive}
\end{align}
is guaranteed only when $\zeta$ is sufficiently larger than $\norm{H}\propto |\Lambda|$, not $\norm{h_{s_1}}$. 
In general, we can only prove the following lemma:
\begin{lemma} \label{positivity_c}
Let $A$ and $B$ be arbitrary operators. 
If $A\succeq 0$ and $B\succeq 0$, 
\begin{align}
e^{A+B+\zeta} - e^A \succeq 0
\end{align}
holds for $\zeta \ge \norm{A}$\footnote{This condition is qualitatively optimal up to a constant factor: if we choose 
$A=\begin{pmatrix}
0 & 0 \\
0 & a \\
\end{pmatrix}
 $
and
$B=\begin{pmatrix}
1 & 1 \\
1 & 1 \\
\end{pmatrix}$, 
the positivity of $e^{A+B+\zeta} - e^A$ is guaranteed only for $c\ge \norm{A}$ in the limit at which $a\to \infty$. 
 }. 
\end{lemma}

\noindent
\textit{Proof.}
$e^{A+B+\zeta} - e^A$ can be rewritten as:
\begin{align}
e^{A+B+\zeta} - e^A=e^{A/2} \br {e^{-A/2+\zeta/2} e^{A+B} e^{-A/2+\zeta/2} -1} e^{A/2}.
\end{align}
Our goal is to prove the positivity of the operator $e^{-A/2+\zeta/2} e^{A+B} e^{-A/2+\zeta/2} -1$. 
First, from $A+B\succeq 0$, for an arbitrary normalized quantum state $\ket{\psi}$ (i.e., $\norm{\ket{\psi}}=1$), we have $\bra{\psi}e^{A+B} \ket{\psi} \ge 1$. 
Second, for an arbitrary normalized quantum state $\ket{\phi}$, the state $\ket{\tilde{\phi}}:=e^{-A/2+\zeta/2}\ket{\phi}$ satisfies $\norm{\ket{\tilde{\phi}}}\ge 1$ due to $-A+\zeta \hat{1} \succeq 0$ from $\zeta \ge \norm{A}$. 
Therefore, combining the two statements above, we have $\bra{\phi} e^{-A/2+\zeta/2} e^{A+B} e^{-A/2+\zeta/2}\ket{\phi} \ge 1$. 
This completes the proof.  $\square$
\\

This implies that we cannot guarantee the positivity of the operator $G_{s_1,s_2}$ by simply shifting the energy origins of $h_{s_1}$ and $h_{s_2}$.
For that reason, we must handle the quantity $(1/Z_\beta)\abs{\tr\br{\Psi_{X,Y} G_{s_1,s_2,\ldots,s_m}}}$ in inequality~\eqref{correlation_up_cluster_1D} without relying on the upper bound $(2/Z_\beta) \norm{ G_{s_1,s_2,\ldots,s_m}}_1$. 
This restricts the use of various mathematical tools for evaluating the upper bounds of the trace norm.

\section{Proof of the main theorem}\label{sec:Proof of the main theorem}

\subsection{Cluster expansion for approximate quantum Gibbs state}
We now consider an approximation of $e^{\beta H}$, denoted by $\tilde{\rho}_\beta$, such that 
\begin{align} 
\norm{ \tilde{\rho}_\beta -e^{\beta H} }_1 \le \epsilon Z_\beta, 
\end{align}
where $\tilde{\rho}_\beta$ is defined using an appropriate polynomial in the interaction terms $\{h_Z\}_{Z\subset \Lambda}$. 
We then apply Lemma~\ref{lem:cluster_exp} to the expansions of $\tilde{\rho}_\beta$ with respect to $\{h_Z\}_{Z\subset \Lambda}$. 
Next, instead of considering $G_{s_1,s_2,\ldots,s_m} $ in Eq.~\eqref{op_G_j_1___s_m}, we analyze
\begin{align} 
\tilde{G}_{s_1,s_2,\ldots,s_m}:= \mathcal{D}_{h_{s_1},\ldots,h_{s_m}}  \tilde{\rho}_\beta , 
\end{align}
(as introduced in \eqref{def_Eq_tilde_G_s_1_s_m} below), which leads to 
\begin{align}
\label{correlation_up_cluster_1D__rederive}
\abs{\Cor_{\rho_\beta}(O_X,O_Y)}  \le 2\epsilon+\abs{\frac{1}{Z_\beta}\Cor_{\tilde{\rho}_\beta}(O_X,O_Y)} 
\le  2\epsilon+\frac{1}{Z_\beta} \tr \br{\Psi_{X,Y} \tilde{G}_{s_1,s_2,\ldots,s_m} } ,
\end{align}
where we used $\norm{\Psi_{X,Y}}\le 2$ [see Eq.~\eqref{def: Psi_X_Y} for the definition of $\Psi_{X,Y}$] and 
\begin{align} 
\abs{\Cor_{\rho_\beta}(O_X,O_Y) -\frac{1}{Z_\beta}\Cor_{\tilde{\rho}_\beta}(O_X,O_Y)} =
\abs{\tr \left[\Psi_{X,Y}  \br{\rho_\beta - \frac{1}{Z_\beta} \tilde{\rho}_\beta}\right]}
\le  \frac{\norm{\Psi_{X,Y}}}{Z_\beta} \norm{e^{\beta H} -  \tilde{\rho}_\beta}_1.
\end{align}


\subsection{Decomposition of the system} \label{sec:Decomposition of the system}

In estimating the operator $G_{s_1,s_2,\ldots,s_m}$ in Eq.~\eqref{op_G_j_1___s_m}, we encountered difficulties, as previously explained: owing to the inequality between the LHS and the RHS of equation \eqref{first_diff_non_comm}, simultaneously removing the zeroth-order terms from the interaction terms is considerably challenging. 
   
\begin{figure}
\centering
\includegraphics[clip, scale=0.47]{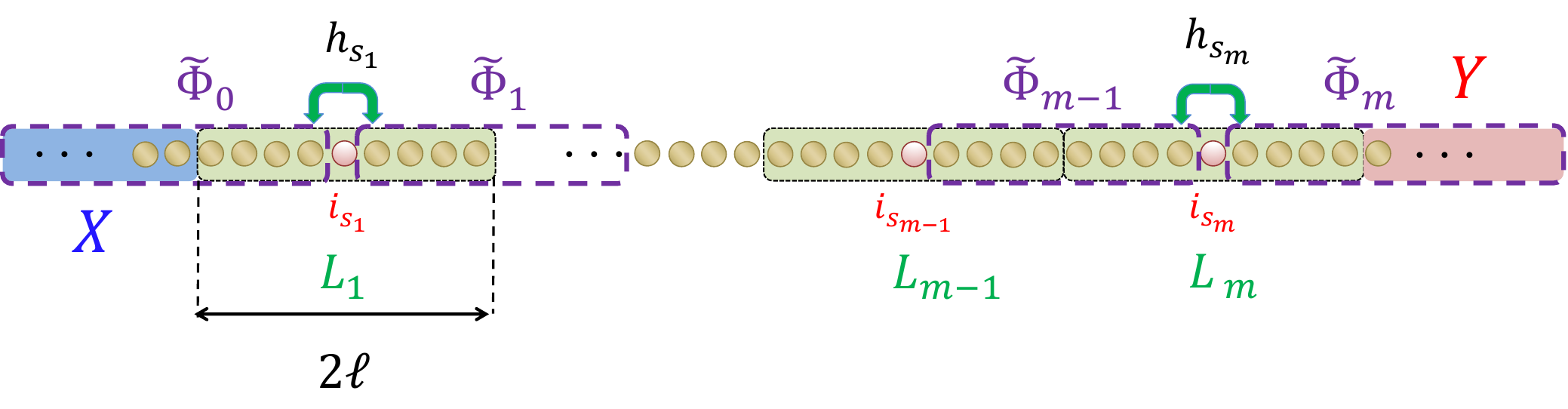}
\caption{Approximating $\{\Phi_j\}_{j=0}^{m+1}$ by $\{\tilde{\Phi}_j\}_{j=0}^{m+1}$, 
the dependence on $h_{s_1}, \ldots, h_{s_m}$ are removed from $\{\tilde{\Phi}_j\}_{j=0}^{m+1}$. 
These approximations allow for a simple analysis to remove the zeroth-order terms for $h_{s_1}, \ldots, h_{s_m}$. 
The center site in the region $L_j$ is denoted by $i_{s_j}$, and the operator $h_{s_j}$ characterizes the interaction between the left half and the right half of the block $L_j$.
}
\label{fig:L_s_Phi_s}
\end{figure}

We decompose the central system $\Lambda\setminus (X\cup Y)$ into $m$ blocks $L_1,L_2,\ldots, L_m$ (Fig.~\ref{fig:L_s_Phi_s}):
\begin{align}
\label{decomp_total_system_L_j}
\Lambda= L_0 \cup L_1 \cup L_2 \cdots L_m \cup L_{m+1} ,  \quad L_0=X, \quad L_{m+1}=Y,
\end{align}
each of length of $2\ell$, i.e., $m \propto R/\ell$.
We choose $\ell$ to satisfy the condition 
\begin{align}
\ell > 6 l_0,  \label{ell>6l/_0_con}
\end{align}
so that Proposition~\ref{lem:belief_error}, under condition~\eqref{r>6l/_0_con}, applies. 
We use $\{i_{s_j}\}_{j=1}^m$ to represent the central sites in $\{L_j\}_{j=1}^m$. Then, using the belief propagation operator, we have 
\begin{align}
e^{\beta H} = \Phi_0 \Phi_1 \cdots \Phi_{m} e^{\beta H_{L_0}} e^{\beta H_{L_1}} \cdots e^{\beta H_{L_m}} e^{\beta H_{L_{m+1}}}\br{\Phi_0 \Phi_1 \cdots \Phi_{m} }^\dagger,
\end{align}
where $\Phi_j$ connects the blocks $L_j$ and $L_{j+1}$. We introduce the notations $\Phi_{0:m}$ and $\tilde{\Phi}_{0:m} $ to define the following products of operators:  
\begin{align}
\Phi_{0:m}:=  \Phi_0 \Phi_1 \cdots \Phi_m, \quad \tilde{\Phi}_{0:m} :=\tilde{\Phi}_0 \tilde{\Phi}_1 \cdots \tilde{\Phi}_m.
\end{align} 

We approximate $\{\Phi_j\}_{j=0}^{m+1}$ by $\{\tilde{\Phi}_j\}_{j=0}^{m+1}$ so that they do not include interaction terms $\{h_{s_j}\}_{j=1}^m$ (as illustrated in Fig.~\ref{fig:L_s_Phi_s}). 
To be more precise, $\tilde{\Phi}_0$, $\{\tilde{\Phi}_j\}_{j=1}^{m}$ and $\tilde{\Phi}_{m+1}$ are
defined by using the subset Hamiltonians $H_{(-\infty,i_{s_1})}$, $H_{(i_{s_j},i_{s_{j+1}})}$ and $H_{(i_{s_m},\infty)}$, respectively. 
From Proposition~\ref{lem:belief_error}, the approximation error between $\Phi_j$ and $\tilde{\Phi}_j$ is given by
\begin{align}
\label{error_phi_tilde_0_m_upp__0_start}
\norm{ \Phi_j -\tilde{\Phi}_j } \le  \mathcal{G}_\beta (\ell ),
\end{align}
where $\mathcal{G}_\beta (\ell)$ is given by Eq.~\eqref{def_mathcal_G}. 
Applying this bound leads to the following lemma:
\begin{lemma} \label{lem:many_BP_approx}
The norm of the difference between $\Phi_{0:m}$ and $\tilde{\Phi}_{0:m}$ is bounded above as follows: 
\begin{align}
\label{error_phi_tilde_0_m_upp__0}
\norm{\Phi_{0:m} -\tilde{\Phi}_{0:m} } \le (m+1) e^{2m\tilde{g} \beta}  \mathcal{G}_\beta (\ell) \le 2 m e^{2m\tilde{g} \beta} \mathcal{G}_\beta (\ell) .
\end{align}
\end{lemma}

\textit{Proof of Lemma~\ref{lem:many_BP_approx}.}
We introduce the notation: 
\begin{align}
\label{error_phi_tilde_0_m_upp__0_proof1}
\Phi_{0:m}^{(s)}:=  \tilde{\Phi}_0 \cdots \tilde{\Phi}_{s-1} \Phi_{s}  \cdots \Phi_{m} .
\end{align}
With this notation, we obtain the following expression: 
\begin{align}
\label{error_phi_tilde_0_m_upp__0_proof2}
\Phi_{0:m} -\tilde{\Phi}_{0:m} = \sum_{s=0}^{m} \br{\Phi_{0:m}^{(s)} - \Phi_{0:m}^{(s+1)}} ,
\end{align}
where we have used $\Phi_{0:m}^{(0)}=\Phi_{0:m}$ and $\Phi_{0:m}^{(m+1)}=\tilde{\Phi}_{0:m}$. 
By applying the upper bound from \eqref{error_phi_tilde_0_m_upp__0_start}, we obtain 
\begin{align}
\label{error_phi_tilde_0_m_upp__0_proof3}
\norm{\Phi_{0:m}^{(s)} - \Phi_{0:m}^{(s+1)}} 
&= \norm{ \tilde{\Phi}_0\cdots \tilde{\Phi}_{s-1}\br{ \Phi_{s}  - \tilde{\Phi}_{s} } \Phi_{s+1} \cdots \Phi_{m} } \notag \\
&\le \norm{ \tilde{\Phi}_0} \cdots \norm{\tilde{\Phi}_{s}} \cdot \mathcal{G}_\beta (\ell ) \cdot \norm{\Phi_{s+1}} \cdots \norm{\Phi_{m} }
\le e^{2m\tilde{g} \beta} \mathcal{G}_\beta (\ell ) ,
\end{align}
where we used the following upper bounds: 
\begin{align}
\label{norm_Phi_j_upper_bound}
\norm{\Phi_j}  \le e^{\beta\norm{h_{s_j} }} \le e^{2\beta\tilde{g}}.
\end{align}
(The same inequalities also apply to $\norm{\tilde{\Phi}_j} $.) Applying inequality~\eqref{error_phi_tilde_0_m_upp__0_proof3} to Eq.~\eqref{error_phi_tilde_0_m_upp__0_proof2} results in the main inequality~\eqref{error_phi_tilde_0_m_upp__0}. This completes the proof. $\square$

{~} \\ 

We denote the sum $H_{L_0}+H_{L_1}+\cdots +H_{L_{m+1}}$ by $H_0$. Then, we have 
\begin{align}
\label{error_phi_tilde_0_m_upp}
\norm{ e^{\beta H}- \tilde{\Phi}_{0:m} e^{\beta H_0}  \tilde{\Phi}_{0 :m}^\dagger}_1 
&= \norm{ e^{\beta H}- \tilde{\Phi}_{0:m} \Phi_{0:m}^{-1} e^{\beta H} \br{\Phi_{0:m}^{\dagger}}^{-1}  \tilde{\Phi}_{0 :m}^\dagger}_1  \notag \\
&= \norm{ \br{1- \tilde{\Phi}_{0:m} \Phi_{0:m}^{-1}}e^{\beta H}  + \tilde{\Phi}_{0:m} \Phi_{0:m}^{-1} e^{\beta H} \brr{1-\br{\Phi_{0:m}^{\dagger}}^{-1}  \tilde{\Phi}_{0 :m}^\dagger}}_1  \notag \\
&\le \br{1+  \norm{ \tilde{\Phi}_{0:m} \Phi_{0:m}^{-1}}} \norm{1- \tilde{\Phi}_{0:m} \Phi_{0:m}^{-1}}  \cdot \norm{e^{\beta H} }_1 \notag \\
&\le (1+  \norm{ \tilde{\Phi}_{0:m}} \cdot \norm{ \Phi_{0:m}^{-1}}) \norm{ \Phi_{0:m}^{-1}}  \cdot \norm{\Phi_{0:m}- \tilde{\Phi}_{0:m}} \tr \br{e^{\beta H} } \notag\\
&\le 4m e^{8(m+1) \beta \tilde{g}} \mathcal{G}_\beta (\ell ) \tr \br{e^{\beta H} } , 
\end{align}
where we used $\norm{\tilde{\Phi}_{0:m}}\le e^{2(m+1)\beta\tilde{g}}$ and $\norm{\Phi_{0:m}^{-1}}\le e^{2(m+1)\beta\tilde{g}}$.
We also introduce the following notation:  
\begin{align}
\label{def_Eq_tilde_G_s_1_s_m}
\tilde{G}_{s_1,s_2,\ldots, s_m}:=\mathcal{D}_{h_{s_1},\ldots,h_{s_m}} \tilde{\Phi}_{0:m} e^{\beta H_{L_0}} e^{\beta H_{L_1}} \cdots e^{\beta H_{L_{m+1}}} \tilde{\Phi}_{0 :m}^\dagger  .
\end{align}
Using inequality~\eqref{correlation_up_cluster_1D__rederive} and taking the limit as $\epsilon \to e^{m \Theta(\beta)} \mathcal{G}_\beta (\ell) $ with error bound \eqref{error_phi_tilde_0_m_upp}, we obtain  
\begin{align}
\label{correlation_up_cluster_1D__rederive_2}
\Cor_{\rho_\beta}(O_X,O_Y) \le e^{m \Theta(\beta)} \mathcal{G}_\beta (\ell) + \frac{1}{Z_\beta} \tr \br{\Psi_{X,Y} \tilde{G}_{s_1,s_2,\ldots,s_m}}.
\end{align} 

Because $\tilde{\Phi}_{0:m}$ does not include the interaction terms $h_{s_1}, \ldots, h_{s_m}$, and Hamiltonians $\{H_{L_j}\}_{j=0}^{m+1}$ commute with one another, Eq.~\eqref{def_Eq_tilde_G_s_1_s_m} simplifies to 
\begin{align}
\label{target_quantityt_Psi_G_s_1}
\tilde{G}_{s_1,s_2,\ldots, s_m}
= \tilde{\Phi}_{0:m} e^{\beta H_{L_0}} \prod_{j=1}^m \brr{ e^{\beta H_{L_j} }- e^{\beta \br{H_{L_j} - h_{s_j}}} } e^{\beta H_{L_{m+1}}} \tilde{\Phi}_{0 :m}^\dagger  .
\end{align}
Now, the remaining and central task is to deduce an upper bound on the second term on the RHS of \eqref{correlation_up_cluster_1D__rederive_2}: 
\begin{align}
\label{target_quantityt_Psi}
\tr\br{\Psi_{X,Y} \tilde{G}_{s_1,s_2,\ldots, s_m} }
&= \tr\br{\Psi_{X,Y} \tilde{\Phi}_{0:m} e^{\beta H_{L_0}} \prod_{j=1}^m \brr{ e^{\beta H_{L_j} }- e^{\beta \br{H_{L_j} - h_{s_j}}} } e^{\beta H_{L_{m+1}}} \tilde{\Phi}_{0 :m} } . 
\end{align}
We state the result here as a subtheorem. The proof of this key subtheorem is postponed to section~\ref{sub_Proof of Subtheorem_main}.
\begin{subtheorem} \label{sbthm:tildeG_upp}
The quantity $\abs{\tr\br{\Psi_{X,Y} \tilde{G}_{s_1,s_2,\ldots, s_m} }} $ has the following upper bound:
\begin{align}
\label{upprer_G_1:q_tr_L_1:q_final_form}
\abs{\tr\br{\Psi_{X,Y} \tilde{G}_{s_1,s_2,\ldots, s_m} }} 
\le&\left( 6 e^{-m e^{-\Theta(\beta)} } +e^{m\Theta(\beta)} \mathcal{G}_\beta (\ell) \right)\tr\br{e^{\beta H }} .
\end{align}
\end{subtheorem}

\subsection{Completing the proof} \label{seec/:Completing the proof}

We are now ready to prove the main theorem.
Since we adopted the interaction-truncated Hamiltonian with interaction length $l_0$, we have  
\begin{align}
\Cor_{\rho_\beta}(O_X,O_Y) \le \Cor_{\rho_{\tc, \beta}}(O_X,O_Y) + \tilde{\gamma} \beta R l_0 \bar{J}(l_0) ,
\end{align}
where we used inequality~\eqref{truncation:error_norm} and introduced the notation $\tilde{\gamma}:=3\gamma^2g$. 
Then, inequality~\eqref{correlation_up_cluster_1D__rederive_2} with Subtheorem~\ref{sbthm:tildeG_upp} gives
\begin{align}
\Cor_{\rho_{\tc, \beta}}(O_X,O_Y) \le e^{m \Theta(\beta)} \mathcal{G}_\beta (\ell) +\left( 6 e^{-m e^{-\Theta(\beta)} } +e^{m\Theta(\beta)} \mathcal{G}_\beta (\ell) \right) .
\end{align}
Combining the above two inequalities yields the following result: 
\begin{align}
\label{clustering_rho_be_taO_X_O_Y_para}
\Cor_{\rho_\beta}(O_X,O_Y) \le e^{m \Theta(\beta)} \min \left( e^{-\ell/[l_0\Theta(\beta)]} , \left(\bar{J}(\ell/3)\right)^{1/\Theta(\beta)} \right) + 6 e^{-m e^{-\Theta(\beta)}}  + \tilde{\gamma} \beta R l_0 \bar{J}(l_0) ,
\end{align}
where we used the expression \eqref{def_mathcal_G} for $\mathcal{G}_\beta(\ell)$.

Now, we introduce the free parameters $\ell$ and $l_0$:
\begin{align}
\ell = \frac{R}{c(R)}, \quad l_0 = \frac{R}{c_0(R)},
\end{align}
where $c(R)$ and $c_0(R)$ are functions of $R$ to be determined below. Owing to condition~\eqref{ell>6l/_0_con}, they satisfy the relation $c(R) < c_0(R)$. We also note that $m\ell$ is proportional to $R$, i.e., $m\ell=c_4R$ for some positive constant $c_4$. Then, inequality~\eqref{clustering_rho_be_taO_X_O_Y_para} simplifies to
\begin{align}
\label{clustering_rho_be_taO_X_O_Y_para2}
&\Cor_{\rho_\beta}(O_X,O_Y) \notag \\
& \le e^{c_4\Theta(\beta) c(R)} \min \left( e^{-\frac{R}{\Theta(\beta\ell_0) c(R)}} , \bar{J}\br{\frac{R}{3c(R)}}^{1/\Theta(\beta) } \right) + 6 e^{-c_4c(R) e^{-\Theta(\beta)}}  + \frac{\tilde{\gamma} \beta R^2}{c_0(R)}  \bar{J}\br{\frac{R}{c_0(R)}} ,
\end{align}
Note that, by the definition of $\Theta(\beta)$, the leading contribution to $e^{-\frac{R}{\Theta(\beta)\ell_0c(R)}}$, which is in the first term on the RHS of \eqref{clustering_rho_be_taO_X_O_Y_para2}, is $e^{-\frac{c_0(R)}{\Theta(\beta)c(R)}}$. It only remains to choose appropriate $c_0(R)$ and $c(R)$ to satisfy inequality~\eqref{main:clustering_rho_subexp} for interactions with sub-exponential decay, and inequality \eqref{main:clustering_rho_slower} for other types of interaction, respectively. This completes the proof of the main theorem.  
 
First, we discuss interactions with sub-exponential decay, i.e., 
\begin{align}
\bar{J}(r) \le e^{-\Theta(r^{\kappa})} .
\end{align}
In this case, we choose 
\begin{align}
c(R) = c_5 c_0(R), 
\end{align}
where $c_5$ represents a positive constant, and 
\begin{align}
 \min \left( e^{-\frac{R}{\Theta(\beta\ell_0) c(R)}} , \bar{J}\br{\frac{R}{3c(R)}}^{1/\Theta(\beta) }\right) 
\le  \bar{J}\left(\frac{R}{3c(R)}\right)^{1/\Theta(\beta)} \le \exp \left(-\frac{1}{\Theta(\beta)} \left(\frac{R}{c(R)}\right)^{\kappa}\right) .
\end{align}
We then impose the following condition on the first term on the RHS of~\eqref{clustering_rho_be_taO_X_O_Y_para}:
\begin{align}
\label{impose_c(R)_condition_1}
 e^{c_4\Theta(\beta) c(R)} \exp \left(-\frac{1}{\Theta(\beta)} \left(\frac{R}{c(R)}\right)^{\kappa}\right)  \le \exp \left(-\frac{1}{2\Theta(\beta)} \left(\frac{R}{c(R)}\right)^{\kappa}\right) ,
\end{align}
which is equivalent to 
\begin{align}
c_4\Theta(\beta)c(R)\le \frac{1}{2\Theta(\beta)}\left(\frac{R}{c(R)}\right)^\kappa.
\end{align}
This holds when we choose
\begin{align}
c(R) =c_6 \frac{R^{\kappa/(\kappa+1)}}{\Theta(\beta)^{2/(\kappa+1)}},
\end{align}
under the condition $0<c_6\le \left(\frac{1}{2c_4}\right)^{1/\kappa+1}$.
With the aforementioned choices for $c_0(R)$ and $c(R)$, we obtain 
\begin{align}
&\Cor_{\rho_\beta}(O_X,O_Y) \notag \\
& \le e^{c_4\Theta(\beta) c(R)} \min \left( e^{-\frac{R}{\Theta(\beta\ell_0) c(R)}} , \bar{J}\br{\frac{R}{3c(R)}}^{1/\Theta(\beta) } \right) + 6 e^{-c_4c(R) e^{-\Theta(\beta)}}  + \frac{\tilde{\gamma} \beta R^2}{c_0(R)}  \bar{J}\br{\frac{R}{c_0(R)}} \notag \\
& \le \exp\left(-\left(\frac{1}{c_6}\right)^\kappa R^\frac{\kappa}{\kappa+1}\Theta(\beta)^\frac{\kappa-1}{\kappa+1}\right)+6 e^{-c_4c_6\frac{R^{\kappa/\kappa+1}}{\Theta(\beta)^{2/\kappa+1}}e^{-\Theta(\beta)}}+\frac{\tilde{\gamma}\beta R^2}{c_0(R)}\,e^{-\Theta\left(\left(\frac{c_5}{c_6}\right)^\kappa R^\frac{\kappa}{\kappa+1}\Theta(\beta)^\frac{2\kappa}{\kappa+1}\right)}.
\end{align}
Therefore, we find that there exists a positive constant $c_1$ such that 
\begin{align}
\label{clustering_rho_be_taO_X_O_Y_subexp}
&\Cor_{\rho_\beta}(O_X,O_Y)  \le c_1 e^{-R^{\kappa/(\kappa+1)}/e^{\Theta(\beta)}},
\end{align}
yielding inequality~\eqref{main:clustering_rho_subexp}.

Next, we discuss decay that is slower than sub-exponential, i.e.,
\begin{align}
\lim_{r\to \infty}e^{-r^\kappa}/\bar{J}(r) = 0 \hspace{.1in}  \forall \kappa>0. 
\end{align}
In this case, we choose 
\begin{align}
 \min \left( e^{-\frac{c_0(R)}{\Theta(\beta) c(R)}} , \bar{J}\left(\frac{R}{3c(R)}\right)^{1/\Theta(\beta) }\right) 
\le \exp\left(-\frac{c_0(R)}{\Theta(\beta) c(R)}\right) . 
\end{align}
Analogous to~\eqref{impose_c(R)_condition_1}, we impose the following condition: 
\begin{align}
 e^{c_4\Theta(\beta) c(R)}  \exp\left(-\frac{c_0(R)}{\Theta(\beta) c(R)}\right)  \le \exp\left(-\frac{c_0(R)}{2\Theta(\beta) c(R)}\right).  
\end{align}
This condition is equivalent to
\begin{align}
c_4 \Theta(\beta) c(R) \le \frac{c_0(R)}{2\Theta(\beta) c(R)},
\end{align}
which is satisfied when $c(R)$ and $c_0(R)$ satisfy the relation
\begin{align}
\label{eq_cr}
c(R) =c_7\,\frac{\sqrt{c_0(R)}}{\Theta(\beta)}
\end{align}
with $0<c_7\le \frac{1}{\sqrt{2c_4}}$.  

Finally, we choose the function $c(R)$ such that it satisfies the inequality 
\begin{align}
e^{-c(R) e^{-\Theta(\beta)}}  \le \bar{J}(R). 
\end{align}
This inequality is satisfied when $c(R)$ satisfies the relation
\begin{align}
c(R) = e^{\Theta(\beta)} \log\left(\bar{J}(R)^{-1}\right).
\end{align}
Applying relation \eqref{eq_cr}, in the form $c_0(R)=\frac{\Theta(\beta)^2 c(R)^2}{c_7^2}$, to this expression for $c(R)$, we deduce that $c_0(R)$ takes the following value:
\begin{align}
c_0(R)=\frac{e^{2\Theta(\beta)} \Theta(\beta)^2 \log^2\left(\bar{J}(R)^{-1}\right)}{c_7^2}.
\end{align} 
For decay that is slower than sub-exponential, the third term on the RHS of inequality~\eqref{clustering_rho_be_taO_X_O_Y_para} yields the dominant contribution. For these values of $c_0(R)$ and $c(R)$, this term can be written explicitly as:
\begin{align}
\frac{\tilde{\gamma}\beta c_7^2 R^2}{\Theta(\beta)^2\, e^{2\Theta(\beta)} \log^2 \left(\bar{J}(R)^{-1}\right)}\,\bar{J}\left(\frac{c_7^2R}{e^{2\Theta(\beta)}\Theta(\beta)^2 \log^2\left(\bar{J}(R)^{-1}\right) }\right). 
\end{align}
Therefore, we deduce that there is a positive constant $c_8$ such that the following inequality holds:
\begin{align}
&\Cor_{\rho_\beta}(O_X,O_Y)  \le  c_8\,\left(\frac{R}{e^{\Theta(\beta)}}\right)^2   \bar{J}\left(\frac{R}{e^{\Theta(\beta)} \log^2\left(\bar{J}(R)^{-1}\right) }\right) .
\end{align}
Selecting a positive constant $c_2$ that satisfies $c_2\ge \frac{c_8}{e^{2\Theta(\beta)}}$ yields inequality~\eqref{main:clustering_rho_slower}. This completes the proof of Theorem~\ref{main_thm_clustering}. $\square$

\section{Proof of Subtheorem~\ref{sbthm:tildeG_upp}}
\label{sub_Proof of Subtheorem_main}

First, we explicitly write the target quantity $\tilde{G}_{s_1,s_2,\ldots, s_m} $ in Eq.~\eqref{target_quantityt_Psi_G_s_1} as follows:
\begin{align}
\label{tilde_m-1_G_1:1_rewrite}
 \tilde{G}_{s_1,s_2,\ldots, s_m}& =\tilde{\Phi}_{0:m}e^{\beta H_{L_0}}  \prod_{j=1}^m \br{e^{\beta H_{L_j}} -e^{\beta (H_{L_j}-h_{s_j})} } e^{\beta H_{L_{m+1}}} \tilde{\Phi}_{0:m}^\dagger   \notag \\
&=\tilde{\Phi}_{0:m}e^{\beta H_{L_0}} \prod_{j=1}^m \br{e^{\beta H_{L_j}(\lambda_j=1)} -e^{\beta H_{L_j}(\lambda_j=0)} }  e^{\beta H_{L_{m+1}}} \tilde{\Phi}_{0:m}^\dagger  \notag \\
&=(-1)^m\sum_{\vec{\lambda}=\{0,1\}^{\otimes m}} (-1)^{\lambda_1+\lambda_2+\cdots+\lambda_m}   \tilde{\Phi}_{0:m}  \exp \br{\beta\sum_{j=0}^{m+1} H_{L_j}(\lambda_j)} \tilde{\Phi}_{0:m}^\dagger ,
\end{align}
where the interaction term $h_{s_j}$ is parametrized by $\lambda_j h_{s_j}$, and the Hamiltonian $H_{L_j} - (1-\lambda_j) h_{s_j}$ is denoted as $H_{L_j}(\lambda_j)$. For $j=0,m+1$, we set $\lambda_0=\lambda_{m+1}=1$, and we set $H_{L_0}(\lambda_0=1)=H_{L_0}$ and $H_{L_{m+1}}(\lambda_{m+1}=1)=H_{L_{m+1}}$.

As defined in Sec.~\ref{sec:Decomposition of the system}, the approximate belief propagation operators $\{\tilde{\Phi}_j\}_{j=0}^{m+1}$ do not include interaction terms $\{h_{s_j}\}_{j=1}^m$ (Fig.~\ref{fig:L_s_Phi_s}). 
Hence, for an arbitrary set $\vec{\lambda}$, we can derive the same inequality as~\eqref{error_phi_tilde_0_m_upp} using Lemma~\ref{lem:many_BP_approx}: 
\begin{align}
\label{tilde_m-1_G_1:1_rewrite_approx}
\norm { \tilde{\Phi}_{0:m}  \exp \br{\beta\sum_{j=0}^{m+1} H_{L_j}(\lambda_j)} \tilde{\Phi}_{0:m}^\dagger
- e^{\beta H(\vec{\lambda})}} 
&\le e^{m\Theta(\beta)} \mathcal{G}_\beta (\ell) \tr\br{e^{\beta H(\vec{\lambda})}}   ,
\end{align}
where $H(\vec{\lambda})$ is the parametrized Hamiltonian as follows:
\begin{align}
H(\vec{\lambda})= H - \sum_{j=1}^m (1-\lambda_j) h_{s_j} . 
\end{align} 
Note that the Golden--Thompson inequality provides 
\begin{align}
\label{tilde_m-1_G_1:1_rewrite_approx/2}
\tr \br{ e^{\beta H(\vec{\lambda})}} \le\tr \br{e^{\beta H } e^{- \beta \sum_{j=1}^m(1-\lambda_j) h_{s_j}}}
&\le \tr \br{e^{\beta H } e^{ \beta \sum_{j=1}^m(1-\lambda_j) \norm{h_{s_j}} }} \notag \\
&\le e^{2\beta m \tilde{g}} \tr \br{e^{\beta H } } ,
\end{align} 
where we used $\norm{h_{s_j}}\le 2\tilde{g}$ as in \eqref{new_h_s_bound}. 
By applying inequalities \eqref{tilde_m-1_G_1:1_rewrite_approx}, \eqref{tilde_m-1_G_1:1_rewrite_approx/2} to Eq.~\eqref{tilde_m-1_G_1:1_rewrite}, we have 
\begin{align}
&\norm{ \tilde{G}_{s_1,s_2,\ldots, s_m}- (-1)^m\sum_{\vec{\lambda}=\{0,1\}^{\otimes m}} (-1)^{\lambda_1+\lambda_2+\cdots+\lambda_m} e^{\beta H(\vec{\lambda})}}_1   \notag \\
&\le 2^m e^{m\Theta(\beta)} \mathcal{G}_\beta (\ell) \tr\br{e^{\beta H}}  \le e^{m\Theta(\beta)} \mathcal{G}_\beta (\ell) \tr\br{e^{\beta H }} ,
\label{norm_tilde_G_s_1___s_m}
\end{align}
which also implies 
\begin{align}
\label{Gamma_L_1_L_m_upper_bound}
&\abs{ \tr\br{\Psi_{X,Y}\tilde{G}_{s_1,s_2,\ldots, s_m} }} \notag \\
& \le  \abs{ \tr\br{\Psi_{X,Y}  (-1)^m\sum_{\vec{\lambda}=\{0,1\}^{\otimes m}} (-1)^{\lambda_1+\lambda_2+\cdots+\lambda_m} e^{\beta H(\vec{\lambda})}}  }+e^{m\Theta(\beta)} \mathcal{G}_\beta (\ell) \tr\br{e^{\beta H }} ,
\end{align}
where we used $\norm{\Psi_{X,Y}}\le 2$.

Next, we consider the operator 
\begin{align}
\label{def:gamma_G}
\Gamma:= (-1)^m\sum_{\vec{\lambda}=\{0,1\}^{\otimes m}} (-1)^{\lambda_1+\lambda_2+\cdots+\lambda_m} e^{\beta H(\vec{\lambda})} .
\end{align}
By applying the operator $\Gamma$ to the inequality~\eqref{Gamma_L_1_L_m_upper_bound}, we have
\begin{align}
\label{Gamma_L_1_L_m_upper_bound__2}
&\abs{ \tr\br{\Psi_{X,Y}\tilde{G}_{s_1,s_2,\ldots, s_m} }} \le  \abs{ \tr\br{\Psi_{X,Y}  \Gamma }}+e^{m\Theta(\beta)} \mathcal{G}_\beta (\ell) \tr\br{e^{\beta H }} .
\end{align}

To analyze the properties of $\Gamma$, we construct an operator to approximate $\Gamma$, and we deduce another bound on $\abs{ \tr\br{\Psi_{X,Y}\tilde{G}_{s_1,s_2,\ldots, s_m} }}$ using the resulting operator. For this purpose, we define $H_{\ge j_0}(\vec{\lambda})$ as 
\begin{align}
H_{\ge j_0}(\vec{\lambda}_{\ge j_0})= H_0 + \sum_{j=j_0}^m  \lambda_j h_{s_j},\quad H_0:=  H (\vec{0}) , \quad H_{\ge 1}(\vec{\lambda})= H(\vec{\lambda}) ,
\end{align}
where we introduced the notation $\vec{\lambda}_{\ge j_0}=\{\lambda_j \}_{j=j_0}^m $. 
We then define the belief propagation operator $\Phi_{s_1}(\vec{\lambda}_{\ge1}) $ as 
\begin{align}
e^{\beta H(\vec{\lambda})} = \Phi_{s_1}(\vec{\lambda}_{\ge1})  e^{\beta \br{H(\vec{\lambda})-\lambda_1 h_{s_1}}} \Phi_{s_1}(\vec{\lambda}_{\ge1})^\dagger 
=\Phi_{s_1}(\vec{\lambda}_{\ge1}) e^{\beta H_{\ge 2}(\vec{\lambda}_{\ge 2})}  \Phi_{s_1}(\vec{\lambda}_{\ge1})^\dagger  .
\end{align}
We note that for $\lambda_1=0$, $\Phi_{s_1}(\vec{\lambda}_{\ge1})$ becomes the identity operator $\hat{1}$.

Next, we define $\Phi_{s_2}(\vec{\lambda}_{\ge2})$ as the belief propagation operator of 
\begin{align}
e^{\beta H_{\ge 2}(\vec{\lambda}_{\ge 2})} = \Phi_{s_2}(\vec{\lambda}_{\ge2}) e^{\beta \br{H_{\ge 2}(\vec{\lambda}_{\ge 2})-\lambda_2 h_{s_2}}} \Phi_{s_2}(\vec{\lambda}_{\ge2}) = \Phi_{s_2}(\vec{\lambda}_{\ge2}) e^{\beta \br{H_{\ge 3}(\vec{\lambda}_{\ge 3})}} \Phi_{s_2}(\vec{\lambda}_{\ge2})  ,
\end{align}
Similarly, we define $\Phi_{s_j}(\vec{\lambda}_{\ge j})$ as the belief propagation operator of 
\begin{align}
\label{def_phi_s_j_lambda}
e^{\beta H_{\ge j}(\vec{\lambda}_{\ge j})}  =\Phi_{s_j}(\vec{\lambda}_{\ge j}) 
e^{\beta H_{\ge j+1}(\vec{\lambda}_{\ge j+1})} \Phi_{s_j}(\vec{\lambda}_{\ge j}), 
\end{align}
(with $\Phi_{s_j}(\vec{\lambda}_{\ge j}) =\hat{1}$ for $\lambda_j=0$), where the Hamiltonian $H_{\ge j+1}(\vec{\lambda}_{\ge j+1})$ includes the interaction terms $h_{s_{j+1}} ,h_{s_{j+2}} , \ldots ,h_{s_m}$. 
Using these operators, we have 
\begin{align}
e^{\beta H(\vec{\lambda})}   = & \Phi_{s_1}(\vec{\lambda}_{\ge 1})  \Phi_{s_2}(\vec{\lambda}_{\ge 2})\cdots  \Phi_{s_m}(\vec{\lambda}_{\ge m}) e^{\beta H (\vec{0}) } \brr{ \Phi_{s_1}(\vec{\lambda}_{\ge 1})  \Phi_{s_2}(\vec{\lambda}_{\ge 2})\cdots  \Phi_{s_m}(\vec{\lambda}_{\ge m}) }^\dagger.
\end{align}

For the belief propagation operator in Eq.~\eqref{def_phi_s_j_lambda}, we can construct the approximate version $\Phi_{s_j}(\lambda_j)$ using only the Hamiltonian $H_{L_j}$, where the interaction term $h_{s_j}$ connect the left half and the right half of the region $L_j$ (as illustrated in Fig.~\ref{fig:L_s_Phi_s}).  
Then, the approximation error is obtained from~\eqref{def_mathcal_G} as follows:
\begin{align}
\norm{\Phi_{s_j}(\vec{\lambda}_{\ge j})  -  \tilde{\Phi}_{s_j,L_j}(\lambda_j)} \le \mathcal{G}_\beta(\ell) ,
\end{align}
where we used the fact that $L_j$ has a width of $2\ell$. Note that $H_{L_j}$ only includes the interaction term $h_{s_j}$ in $\{h_{s_j}\}_{j=1}^m$; therefore, the operator $\tilde{\Phi}_{s_j,L_j}(\vec{\lambda}_{\ge j})$ depends only on $\lambda_j$. 
By combining the above upper bound with Lemma~\ref{lem:many_BP_approx}, we obtain
\begin{align}
\label{approx_e_b_eta_H_L_1:q}
&\norm{ e^{\beta H(\vec{\lambda})}  - 
  \tilde{\Phi}_{s_1,L_1}(\lambda_1)   \tilde{\Phi}_{s_2,L_2}(\lambda_2)\cdots  \tilde{\Phi}_{s_m,L_m}(\lambda_m) e^{\beta H (\vec{0}) } \brr{ \tilde{\Phi}_{s_1,L_1}(\lambda_1)   \tilde{\Phi}_{s_2,L_2}(\lambda_2)\cdots  \tilde{\Phi}_{s_m,L_m}(\lambda_m)  }^\dagger}_1 \notag \\
 \le &  e^{m\Theta(\beta)}\mathcal{G}_\beta(\ell)  \tr \br{e^{\beta H(\vec{\lambda})} } \le  e^{m\Theta(\beta)}\mathcal{G}_\beta(\ell)  \tr \br{e^{\beta H  }} ,
\end{align}
where we used the relation \eqref{tilde_m-1_G_1:1_rewrite_approx/2} in the last inequality.

Now, we define the operator $\tilde{\Gamma}$ as
\begin{align}
\tilde{\Gamma}:= (-1)^m\sum_{\vec{\lambda}=\{0,1\}^{\otimes m}} (-1)^{\lambda_1+\lambda_2+\cdots+\lambda_m}  \tilde{\Phi}_{s_1,L_1}(\lambda_1) \cdots  \tilde{\Phi}_{s_m,L_m}(\lambda_m) e^{\beta H (\vec{0}) } \brr{ \tilde{\Phi}_{s_1,L_1}(\lambda_1)  \cdots  \tilde{\Phi}_{s_m,L_m}(\lambda_m)  }^\dagger 
\end{align}
to approximate $\Gamma$ in Eq.~\eqref{def:gamma_G}. Then, from~\eqref{approx_e_b_eta_H_L_1:q}, we obtain the following inequality:  
\begin{align}
\label{approx_Gamma_tilde_Gamma}
&\norm{\Gamma - \tilde{\Gamma}}_1 \le  e^{m\Theta(\beta)}\mathcal{G}_\beta(\ell)  \tr \br{e^{\beta H  }}  ,
\end{align}
where we used an inequality similar to~\eqref{norm_tilde_G_s_1___s_m}.
By combining this inequality with the upper bound in \eqref{Gamma_L_1_L_m_upper_bound__2}, we obtain 
\begin{align}
\label{upprer_G_1:q_tr_L_1:q}
\abs{ \tr\br{\Psi_{X,Y} \tilde{G}_{s_1,s_2,\ldots, s_m} }} 
\le&\abs{ \tr\br{\Psi_{X,Y}   \tilde{\Gamma} }} +e^{m\Theta(\beta)} \mathcal{G}_\beta (\ell) \tr\br{e^{\beta H }} .
\end{align}
The only remaining task is to estimate the upper bound of $\abs{ \tr\br{\Psi_{X,Y}   \tilde{\Gamma} }} $. 

We note that $\tilde{\Phi}_{s_j,L_j}(\lambda_j)$ is an approximate belief propagation operator supported on $L_j$. 
Hence, we can guarantee that $[\tilde{\Phi}_{s_j,L_j}(\lambda_j),\tilde{\Phi}_{s_{j'},L_{j'}}(\lambda_{j'}) ]=0$ and that $[\tilde{\Phi}_{s_j,L_j}(\lambda_j), \Psi_{X,Y}]=0$ as well. We remind the reader that the subsets $X$ and $Y$ corresponds to $L_0$ and $L_{m+1}$, respectively [see Eq.~\eqref{decomp_total_system_L_j}]. 
Therefore, we obtain the following decomposition, which is analogous to the classical case~\eqref{trace_norm_G_replce}: 
\begin{align}
&\tr\br{\Psi_{X,Y}   \tilde{\Gamma} }   \notag \\
&=  (-1)^m\sum_{\vec{\lambda}=\{0,1\}^{\otimes m}} (-1)^{\lambda_1+\lambda_2+\cdots+\lambda_m}
\tr\br{\Psi_{X,Y}  \tilde{\Phi}_{s_1,L_1}(\lambda_1) \cdots  \tilde{\Phi}_{s_m,L_m}(\lambda_m) e^{\beta H (\vec{0}) } \brr{ \tilde{\Phi}_{s_1,L_1}(\lambda_1)  \cdots  \tilde{\Phi}_{s_m,L_m}(\lambda_m)  }^\dagger } \notag \\
&=  \tr \brr{\Psi_{X,Y}   \prod_{j=1}^{m} \br{ \tilde{\Phi}_{s_j,L_j}^\dagger  \tilde{\Phi}_{s_j,L_j} -1} e^{\beta H (\vec{0}) } } 
  \notag \\
&=\tr \brr{(\Psi_{X,Y}+2)^{1/2}   \prod_{j=1}^{m} \br{ \tilde{\Phi}_{s_j,L_j}^\dagger  \tilde{\Phi}_{s_j,L_j} -1} e^{\beta H (\vec{0}) } (\Psi_{X,Y}+2)^{1/2}} 
-2 \tr \brr{  \prod_{j=1}^{m} \br{ \tilde{\Phi}_{s_j,L_j}^\dagger  \tilde{\Phi}_{s_j,L_j} -1} e^{\beta H (\vec{0}) }} ,
\label{trace_tulde_Gamma_upp}
\end{align}
where we used $\tilde{\Phi}_{s_j,L_j}(0)=\hat{1}$ and denoted $\tilde{\Phi}_{s_j,L_j}(1)$ as $\tilde{\Phi}_{s_j,L_j}$ for simplicity. The last inequality follows directly from using $(\Psi_{X,Y}+2)^{1/2} (\Psi_{X,Y}+2)^{1/2} - 2 = \Psi_{X,Y}$.
Note that $\Psi_{X,Y}+2$ is positive semidefinite because $\norm{\Psi_{X,Y}}\le 2$.   
To compute $\abs{\tr\br{\Psi_{X,Y}   \tilde{\Gamma} } }$, we generally consider
\begin{align}
&\tr \br{O_{X,Y} \tr_{L_{1:m}}\brr{ \prod_{j=1}^{m} \br{ \tilde{\Phi}_{s_j,L_j}^\dagger  \tilde{\Phi}_{s_j,L_j} -1} e^{\beta H (\vec{0}) }}O_{X,Y}  },
\label{eq:tilde_phi_tr_L_2:q-1//000}
\end{align}
where $O_{X,Y}$ is an arbitrary positive semidefinite Hermitian operator supported on $X\cup Y$.
We then calculate
\begin{align}
\Bra{\tilde{\psi}_{X,Y}} \tr_{L_{1:m}} \brr{ \prod_{j=1}^{m} \br{ \tilde{\Phi}_{s_j,L_j}^\dagger  \tilde{\Phi}_{s_j,L_j} -1} e^{\beta H (\vec{0}) }}
\Ket{\tilde{\psi}_{X,Y}} ,
\label{eq:tilde_phi_tr_L_2:q-1//}
\end{align}
where we used the following notation:
\begin{align}
\ket{\tilde{\psi}_{X,Y}}:=O_{X,Y}  \ket{\psi_{X,Y}} , 
\end{align}
and $\ket{\psi_{X,Y}}$ represents an arbitrary quantum state supported on the subset $X\cup Y$.

We then prove the following lemma, which we will use to evaluate Eq.~\eqref{eq:tilde_phi_tr_L_2:q-1//}. 
\begin{lemma} \label{lem:W_j_W_j-1_ineq}
Let $\{W_j\}_{j=1}^m$ be mutually commuting semi-definite operators (i.e., $W_j\succeq 0$, and $[W_i, W_j]=0$) supported on $X\subset\Lambda$. Then, the following inequality holds for an arbitrary quantum state $\ket{\psi_{X^\co}}$ supported on the subset $X^\co$:  
\begin{align}
 \label{lem:W_j_W_j-1_ineq/main_ineq}
\bra{\psi_{X^\co}} \tr_X \br{\rho \prod_{j=1}^m W_j } \ket{\psi_{X^\co}}\le 
\bra{\psi_{X^\co}} \tr_X \brr{\rho \prod_{j=1}^m (W_j+1) } \ket{\psi_{X^\co}} \cdot \prod_{j=1}^m \frac{\norm{W_j}}{\norm{W_j}+1} ,
\end{align}
where $\rho$ represents an arbitrary positive semi-definite operator.
\end{lemma}

\textit{Proof.}
We first introduce notations: 
\begin{align}
\rho_X:= \bra{\psi_{X^\co}} \rho \ket{\psi_{X^\co}} ,\quad W_j:=\sum_x w_{j,x} \ket{x}\bra{x} ,
\end{align}
where the latter one is the spectral decomposition of $W_j$ using the simultaneous eigenstates $\{\ket{x}\}_x$ of $\{W_j\}_{j=1}^m$
since we have assumed $[W_j,W_{j'}]=0$ for any pair $j,j'$.   
Because $\{W_j\}_{j=1}^m$ is positive semi-definite, the relation $w_{j,x} \ge 0$ holds $\forall j,x$. 
Therefore, we have the following equalities:
\begin{align}
\label{rel_trace}
\bra{\psi_{X^\co}} \tr_X \br{\rho \prod_{j=1}^m W_j } \ket{\psi_{X^\co}}  = 
\tr_X  \br{\rho_X \prod_{j=1}^m W_j } = \sum_x \bra{x} \rho_X \ket{x}\prod_{j=1}^m w_{j,x} .
\end{align} 
In the same manner, we have 
\begin{align}
\label{rel_trace__2}
\bra{\psi_{X^\co}} \tr_X \brr{\rho \prod_{j=1}^m (W_j+1) } \ket{\psi_{X^\co}}  = \sum_x \bra{x} \rho_X \ket{x} \prod_{j=1}^m \br{ w_{j,x} +1}.
\end{align} 
Denoting $ \bra{x} \rho_X \ket{x}$ by $\rho_x$, we obtain the equality: 
\begin{align}
\bra{\psi_{X^\co}} \tr_X \br{\rho \prod_{j=1}^m W_j } \ket{\psi_{X^\co}}  = 
\bra{\psi_{X^\co}} \tr_X \brr{\rho \prod_{j=1}^m (W_j+1) } \ket{\psi_{X^\co}}   
\frac{ \sum_x  \rho_x \prod_{j=1}^mw_{j,x} }{ \sum_x \rho_x \prod_{j=1}^m \br{ w_{j,x} +1}}.
\end{align} 

Next, we define 
\begin{align}
\tilde{W}_{j_0}:= \sum_x \rho_x \prod_{j=1}^{j_0} \br{ w_{j,x} +1}  \prod_{j=j_0+1}^{m} w_{j,x} ,
\end{align} 
where $\tilde{W}_0= \sum_x \rho_x \prod_{j=1}^{m} w_{j,x}$ and $\tilde{W}_m= \sum_x \rho_x \prod_{j=1}^{m} (w_{j,x}+1)$.
We then consider 
\begin{align}
\label{comparison_W_j_x+1_W_j_x}
\frac{ \sum_x  \rho_x \prod_{j=1}^mw_{j,x} }{ \sum_x \rho_x \prod_{j=1}^m \br{ w_{j,x} +1}} 
= \frac{\tilde{W}_0}{\tilde{W}_m} = 
\frac{\tilde{W}_0}{\tilde{W}_1} \cdot \frac{\tilde{W}_1}{\tilde{W}_2}\cdot  \frac{\tilde{W}_2}{\tilde{W}_3}  \cdots   \frac{\tilde{W}_{m-1}}{\tilde{W}_m}  .
\end{align} 
For $j_0 \in [1,m]$, we have 
\begin{align}
\frac{\tilde{W}_{j_0-1}}{\tilde{W}_{j_0}}  
&= \frac{ \sum_x \rho_x \prod_{j=1}^{j_0-1} \br{ w_{j,x} +1}  w_{j_0,x} \prod_{j=j_0+1}^{m} w_{j,x} }{ \sum_x \rho_x \prod_{j=1}^{j_0-1} \br{ w_{j,x} +1} ( w_{j_0,x}+1) \prod_{j=j_0+1}^{m} w_{j,x} } \notag \\
&= \frac{ \sum_x \rho_x w_{j_0,x} Q_{j_0,x}}{ \sum_x \rho_x (w_{j_0,x}+1) Q_{j_0,x} },
\end{align} 
where we introduced the notation $Q_{j_0,x}$ as:
\begin{align}
Q_{j_0,x}:=  \prod_{j=1}^{j_0-1} \br{ w_{j,x} +1} \prod_{j=j_0+1}^{m} w_{j,x}.
\end{align} 
We then deduce the following inequality: 
\begin{align}
\frac{d}{d_{w_{j_0,x}}}  \frac{ \sum_x \rho_x w_{j_0,x} Q_{j_0,x}}{ \sum_x \rho_x (w_{j_0,x}+1) Q_{j_0,x} }
&= \frac{\rho_x Q_{j_0,x}}{ \sum_x \rho_x (w_{j_0,x}+1) Q_{j_0,x} }
- \frac{ \sum_x \rho_x w_{j_0,x} Q_{j_0,x}}{ \sum_x \rho_x (w_{j_0,x}+1) Q_{j_0,x} }
\cdot  \frac{ \rho_x  Q_{j_0,x}}{ \sum_x \rho_x (w_{j_0,x}+1) Q_{j_0,x} } \notag \\
&=  \frac{\rho_x Q_{j_0,x}}{ \sum_x \rho_x (w_{j_0,x}+1) Q_{j_0,x} }\br{1- \frac{ \sum_x \rho_x w_{j_0,x} Q_{j_0,x}}{ \sum_x \rho_x (w_{j_0,x}+1) Q_{j_0,x} }}\ge 0,
\end{align} 
where we used the positivity of $\{w_{j,x}\}_{j,x}$, implying $Q_{j_0,x}\ge 0$.
Because $w_{j_0,x} \le \norm{W_{j_0}}$, we can derive the following inequality  
\begin{align}
\label{ineq_W_j_0_j_0_+1}
\frac{\tilde{W}_{j_0-1}}{\tilde{W}_{j_0}}  
\le \frac{ \sum_x \rho_x \norm{W_{j_0}}Q_{j_0,x}}{ \sum_x \rho_x (\norm{W_{j_0}}+1) Q_{j_0,x} }  
= \frac{ \norm{W_{j_0}}}{ \norm{W_{j_0}}+1}   . 
\end{align} 
Application of the above upper bound to Eq.~\eqref{comparison_W_j_x+1_W_j_x} provides the desired inequality~\eqref{lem:W_j_W_j-1_ineq/main_ineq}, completing the proof. 
 $\square$

{~} \\

We apply Lemma~\ref{lem:W_j_W_j-1_ineq} to Eq.~\eqref{eq:tilde_phi_tr_L_2:q-1//} by substituting $L_{1:m}$ for $X$, $\frac{\ket{\tilde{\psi}_{X,Y}} }{\norm{O_{X,Y}\ket{\psi_{X,Y}} }}$ for $\ket{\psi_X}$, and $\tilde{\Phi}_{s_j,L_j}^\dagger  \tilde{\Phi}_{s_j,L_j} -1$ for $W_j$.  
Here, from Eq.~\eqref{def_tilde_Phi_phi}, the positivity of the operator $\tilde{\Phi}_{s_j,L_j}^\dagger  \tilde{\Phi}_{s_j,L_j} -1$ is ensured by the positivity of the operator $h_s$. We note that $h_Z\succeq 0$ from Eq.~\eqref{h_Z_positive}. 
We then obtain 
\begin{align}
& \Bra{\tilde{\psi}_{X,Y}}\tr_{L_{1:m} }  \brr{ \prod_{j=1}^{m} \br{ \tilde{\Phi}_{s_j,L_j}^\dagger  \tilde{\Phi}_{s_j,L_j} -1} e^{\beta H (\vec{0}) }}
 \Ket{\tilde{\psi}_{X,Y}} \notag \\
&=\norm{O_{X,Y}\ket{\psi_{X,Y}} }^2 \frac{ \Bra{\tilde{\psi}_{X,Y}}}{\norm{O_{X,Y}\ket{\psi_{X,Y}} }}\tr_{L_{1:m} }  \brr{ \prod_{j=1}^{m} \br{ \tilde{\Phi}_{s_j,L_j}^\dagger  \tilde{\Phi}_{s_j,L_j} -1} e^{\beta H (\vec{0}) }}
 \frac{\Ket{\tilde{\psi}_{X,Y}} }{\norm{O_{X,Y}\ket{\psi_{X,Y}} }}\notag \\
&\le\Bra{\tilde{\psi}_{X,Y}} \tr_{L_{1:m} }  \brr{ \prod_{j=1}^{m}  \tilde{\Phi}_{s_j,L_j}^\dagger  \tilde{\Phi}_{s_j,L_j}  e^{\beta H (\vec{0}) }}
\Ket{\tilde{\psi}_{X,Y}} \cdot  \prod_{j=1}^{m} \frac{\norm{\tilde{\Phi}_{s_j,L_j}^\dagger  \tilde{\Phi}_{s_j,L_j} }-1}{\norm{\tilde{\Phi}_{s_j,L_j}^\dagger  \tilde{\Phi}_{s_j,L_j}}} \notag \\
&= \Bra{\psi_{X,Y}}O_{X,Y}
\tr_{L_{1:m} } \brr{ \tilde{\Phi}_{s_1,L_1} \cdots \tilde{\Phi}_{s_m,L_m} e^{\beta H (\vec{0}) }  \br{\tilde{\Phi}_{s_1,L_1} \cdots \tilde{\Phi}_{s_m,L_m}}^\dagger } O_{X,Y}
\Ket{\psi_{X,Y}} \prod_{j=1}^{m} \frac{\norm{\tilde{\Phi}_{s_j,L_j}^\dagger  \tilde{\Phi}_{s_j,L_j} }-1}{\norm{\tilde{\Phi}_{s_j,L_j}^\dagger  \tilde{\Phi}_{s_j,L_j}}}.
\label{upper_bound_for_trace_tulde_Gamma_upp}
\end{align}
Recall that $\tilde{\Phi}_{s_j,L_j}(1)$ is denoted simply as $\tilde{\Phi}_{s_j,L_j}$ for brevity, as in \eqref{trace_tulde_Gamma_upp}. 
Here, we denote the spectral decomposition of the Hilbert space on the subset $X,Y$ by $\hat{1}_{X,Y}$, i.e., $\hat{1}_{X,Y} :=  \sum_{x}  \Ket{x_{X,Y}}\Bra{x_{X,Y}}$. 
By applying the upper bound~\eqref{upper_bound_for_trace_tulde_Gamma_upp} to Eq.~\eqref{eq:tilde_phi_tr_L_2:q-1//000}, we have 
\begin{align}
&\tr \br{O_{X,Y} \brr{\prod_{j=1}^{m} \br{ \tilde{\Phi}_{s_j,L_j}^\dagger  \tilde{\Phi}_{s_j,L_j} -1} e^{\beta H (\vec{0}) }} O_{X,Y}} \notag \\
&= \sum_{x} 
 \Bra{x_{X,Y}}O_{X,Y} 
 \tr_{L_{1:m} }  \brr{   \prod_{j=1}^{m} \br{ \tilde{\Phi}_{s_j,L_j}^\dagger  \tilde{\Phi}_{s_j,L_j} -1} e^{\beta H (\vec{0}) } } O_{X,Y} \Ket{x_{X,Y}} 
 \notag \\
&\le \prod_{j=1}^{m} \frac{\norm{\tilde{\Phi}_{s_j,L_j}^\dagger  \tilde{\Phi}_{s_j,L_j} }-1}{\norm{\tilde{\Phi}_{s_j,L_j}^\dagger  \tilde{\Phi}_{s_j,L_j}}}
 \sum_{x}  \Bra{x_{X,Y}}O_{X,Y}
\tr_{L_{1:m} } \brr{ \tilde{\Phi}_{s_1,L_1} \cdots \tilde{\Phi}_{s_m,L_m} e^{\beta H (\vec{0}) }  \br{\tilde{\Phi}_{s_1,L_1} \cdots \tilde{\Phi}_{s_m,L_m}}^\dagger }  O_{X,Y} \Ket{x_{X,Y}}  \notag \\
&\le  \prod_{j=1}^{m} \frac{\norm{\tilde{\Phi}_{s_j,L_j}^\dagger  \tilde{\Phi}_{s_j,L_j} }-1}{\norm{\tilde{\Phi}_{s_j,L_j}^\dagger  \tilde{\Phi}_{s_j,L_j}}}
\tr \brr{O_{X,Y} \tilde{\Phi}_{s_1,L_1} \cdots \tilde{\Phi}_{s_m,L_m} e^{\beta H (\vec{0}) }  \br{\tilde{\Phi}_{s_1,L_1} \cdots \tilde{\Phi}_{s_m,L_m}}^\dagger   O_{X,Y}   }
\notag \\
&\le  \prod_{j=1}^{m} \frac{\norm{\tilde{\Phi}_{s_j,L_j}^\dagger  \tilde{\Phi}_{s_j,L_j} }-1}{\norm{\tilde{\Phi}_{s_j,L_j}^\dagger  \tilde{\Phi}_{s_j,L_j}}}
\brr{ \tr \br{O_{X,Y} e^{\beta H }  O_{X,Y}} + \norm{O_{X,Y}}^2 e^{m\Theta(\beta)}\mathcal{G}_\beta(\ell) \tr\br{e^{\beta H}}} \notag \\
&\le  \norm{O_{X,Y}}^2 \brr{ 1+e^{m\Theta(\beta)}\mathcal{G}_\beta(\ell)}\tr\br{e^{\beta H}}  \prod_{j=1}^{m} \frac{\norm{\tilde{\Phi}_{s_j,L_j}^\dagger  \tilde{\Phi}_{s_j,L_j} }-1}{\norm{\tilde{\Phi}_{s_j,L_j}^\dagger  \tilde{\Phi}_{s_j,L_j}}} ,
\end{align}
where the inequality~\eqref{approx_e_b_eta_H_L_1:q} was used in the third comparison.
By applying this result to $\abs{\tr\br{\Psi_{X,Y}   \tilde{\Gamma} }}$ in Eq.~\eqref{trace_tulde_Gamma_upp}, we  deduce
\begin{align}
\abs{\tr\br{\Psi_{X,Y}   \tilde{\Gamma} } }  \le \br{\norm{\Psi_{X,Y}}+4} \brr{ 1+e^{m\Theta(\beta)}\mathcal{G}_\beta(\ell)}\tr\br{e^{\beta H}}  \prod_{j=1}^{m} \frac{\norm{\tilde{\Phi}_{s_j,L_j}^\dagger  \tilde{\Phi}_{s_j,L_j} }-1}{\norm{\tilde{\Phi}_{s_j,L_j}^\dagger  \tilde{\Phi}_{s_j,L_j}}} . 
\end{align}

Finally, from inequality~\eqref{norm_Phi_j_upper_bound}, we have  
\begin{align}
\norm{\tilde{\Phi}_{s_j,L_j}^\dagger  \tilde{\Phi}_{s_j,L_j} } \le e^{\Theta(\beta)} , 
\end{align}
and thus, we obtain 
\begin{align}
 \prod_{j=1}^{m} \frac{\norm{\tilde{\Phi}_{s_j,L_j}^\dagger  \tilde{\Phi}_{s_j,L_j} }-1}{\norm{\tilde{\Phi}_{s_j,L_j}^\dagger  \tilde{\Phi}_{s_j,L_j}}} 
 \le  \prod_{j=1}^{m} \br{1- e^{-\Theta(\beta)} }  \le  e^{-m e^{-\Theta(\beta)} },
\end{align}
where we used $(1+x)^m \le e^{mx}$, which holds for real $x$ with $|x|\le 1$ and $m\in \mathbb{N}$. 
We therefore arrive at the inequality 
\begin{align}
\label{inq_tilde_Gamma_final}
&\abs{ \tr \br{\Psi_{X,Y} \tilde{\Gamma}}  }   \le 6 \brr{1+e^{m\Theta(\beta)}\mathcal{G}_\beta(\ell)}  e^{-m e^{-\Theta(\beta)} }  \tr \br{e^{\beta H  }} ,
\end{align}
where we used $\norm{\Psi_{X,Y}}\le 2$. 
Combining inequalities~\eqref{inq_tilde_Gamma_final} and~\eqref{upprer_G_1:q_tr_L_1:q} proves the main inequality~\eqref{upprer_G_1:q_tr_L_1:q_final_form}.
This completes the proof. $\square$

\section{Concluding remarks}
\label{sec5}

In this paper, we have demonstrated a clustering theorem that applies to systems beyond those with finite-range or exponentially decaying interactions. The core findings and significant observations are encapsulated in Theorem~\ref{main_thm_clustering}.  Our methodology integrates the quantum belief propagation technique, combined with a specific decomposition approach outlined in Section~\ref{sec:Decomposition of the system}. 
This approach enables us to replicate analyses typical of classical one-dimensional systems, effectively circumventing the cluster expansion divergence (see Section~\ref{1D cases: commuting Hamiltonian}). 
Our findings conclusively address several aspects of the clustering theorem within one-dimensional quantum systems:
i) the dependence of correlation length on temperature,
ii) the behavior of correlation decay rates for interactions decaying slower than subexponential forms.

Despite these advancements, our study unveils several unresolved issues. 
Among the most critical is improving the correlation decay rate in one-dimensional systems with subexponential (or faster) interaction decay. 
This challenge primarily stems from the approximation used in the inequality~\eqref{error_phi_tilde_0_m_upp}. 
To minimize the error in the approximate belief propagation, i.e., $\mathcal{G}_\beta(\ell)$, a larger block size $\ell$ is necessary. 
Yet, this results in fewer blocks ($m \propto R/\ell$), leading to a less strict decay rate of $e^{-m/e^{\Theta(\beta)}}$ as per equation~\eqref{upprer_G_1:q_tr_L_1:q_final_form}. 
At this stage, the improvement is not a straightforward task, and addressing this issue may require fundamentally novel analytical methods.

One potential strategy for overcoming this challenge involves analyzing the zero-free region of the partition function, as discussed in~\cite{Mehdi2020}. This entails to find a complex number $z$ such that the partition function satisfies $$Z_{\beta + i z} = \tr \br{e^{(\beta+iz)H}}\neq0$$ for any $|z| \le \Delta$, with $\Delta>0$. By proving a zero-free region for the partition function for $\Delta = \Omega(1)$, we can prove the exponential clustering theorem from~\cite[Theorem~10 therein]{Mehdi2020}. Achieving this would leverage a bootstrapping method: initially applying the current `loose' clustering theorem to establish the zero-free region, then applying this result to refine the clustering of correlation based on Ref.~\cite{Mehdi2020}. 
This approach is anticipated to be viable if the quasi-locality $\mathcal{G}_\beta(r)$ decays exponentially with distance $r$, applicable in scenarios of interactions decaying (faster than) exponentially.

Furthermore, analyzing the complex zeros of the partition function reveals additional implications.  The current clustering theorem does not exclude the possibility of a phase transition in systems with power-law decaying interaction since the correlation decay is polynomial in both critical and non-critical phases. A zero-free region for $|z| \le \Delta=\Omega(1)$ leads to the partition function's analyticity in the thermodynamic limit, suggesting the absence of phase transition in long-range interacting systems. 
So far, there are no quantum analogs to the classical no-go theorem~\cite{Dobrushin1973} for the phase transition in long-range interacting systems. 
We speculate that the critical condition for the power-law decay rate $\alpha$ might also be $2$ in quantum scenarios, yet this remains a significant open question. We hope that our new analytical techniques, including the quantum belief propagation, will pave the way for resolving further open questions regarding the 1D long-range interacting systems.

\section{Acknowledgements}

Y. K. and T. K. acknowledge the Hakubi projects of RIKEN.
T. K. was supported by JST PRESTO
(Grant No. JPMJPR2116), ERATO (Grant No. JPMJER2302),
and JSPS Grants-in-Aid for Scientific Research
(No. JP23H01099, JP24H00071), Japan. 
Y. K. was supported by the JSPS Grant-in-Aid for Scientific Research (No. JP24K06909), Japan. We are grateful to Nikhil Srivastava, Angela Capel, and Antonio Per\'ez Hern\'andez for helpful discussions and comments on related topics.

\section{Declarations}
\subsection{Conflict of interest}
The authors declare that there is no conflict of interest. Data sharing not applicable to
this article as no datasets were generated or analysed during the current study.

\def\bibsection{\section*{References}} 
\providecommand{\noopsort}[1]{}\providecommand{\singleletter}[1]{#1}%
%


\end{document}